\documentclass[a4paper,fleqn,usenatbib]{mnras}   %{mn2e}

\usepackage{newtxtext,newtxmath}
\usepackage[T1]{fontenc}
\usepackage{ae,aecompl}

\usepackage{graphicx}	% Including figure files
\usepackage{amsmath}	% Advanced maths commands
\usepackage{amssymb}	% Extra maths symbols
\usepackage{ulem}

\newcommand{\be}{\begin{equation}}
\newcommand{\ee}{\end{equation}}
\newcommand{\ba}{\begin{eqnarray}}
\newcommand{\ea}{\end{eqnarray}}
\newcommand{\hMpc}{{\ifmmode{h^{-1}\,{\rm Mpc}}
\else{$h^{-1}$Mpc}\fi}}
\newcommand{\bk}{\boldsymbol{k}}

\newcommand{\bn}{\boldsymbol{n}}
\newcommand{\bx}{\boldsymbol{x}}
\newcommand{\by}{\boldsymbol{y}}

\newcommand{\bp}{\boldsymbol{p}}
\newcommand{\bq}{\boldsymbol{q}}
\newcommand{\bu}{\boldsymbol{u}}
\newcommand{\bw}{\boldsymbol{w}}
\newcommand{\bs}{\boldsymbol{s}}
\newcommand{\bee}{\boldsymbol{e}}
\newcommand{\ngal}{n_{\rm g}}
\newcommand{\br}{\boldsymbol{r}}
\title[Cosmology from the bispectrum]{Cosmological information in the redshift-space bispectrum} 
\author[Yankelevich \& Porciani]{
Victoria Yankelevich\thanks{Member of the International Max Planck Research School (IMPRS) for Astronomy and Astrophysics at the Universities of Bonn and Cologne}\thanks{E-mail: vyankelevich@astro.uni-bonn.de} \& 
Cristiano Porciani\thanks{E-mail: 
porciani@astro.uni-bonn.de}
\\
Argelander-Institut f\"ur Astronomie, University of Bonn, Auf dem H\"ugel 71, D-53121 Bonn, Germany
}

\date{Accepted XXX. Received YYY; in original form ZZZ}

\pubyear{2018}

\begin{document}
\label{firstpage}
\pagerange{\pageref{firstpage}--\pageref{lastpage}}
\maketitle

\begin{abstract}
We use the Fisher-matrix formalism to investigate whether the galaxy bispectrum in redshift space, $B$, contains additional cosmological information with respect to the power spectrum, $P$. 
We focus on a \textit{Euclid}-like survey and consider cosmological models dominated by dark energy and cold dark matter with Gaussian primordial perturbations.
After discussing the phenomenology of redshift-space distortions for
the bispectrum, we derive an expression for the cross-covariance
between $B$ and $P$ at leading order in perturbation theory. Our equation generalizes previous results
that did not consider binning in the orientation of wavevector triangles with respect to the line of sight. By considering Fourier modes with wavenumber $k<0.15 \,h$ Mpc$^{-1}$, we find that 
$B$ and $P$ set similar constraints on the cosmological parameters. 
Generally, error bars moderately improve when the two probes are combined together.
For instance, the joint 68.3 per cent credible region for the parameters that describe a dynamical dark-energy equation of state
shrinks by a factor of 2.6 with respect to only using the power spectrum. 
Regrettably, this improvement is cancelled out when the clustering analysis is combined with priors based on current studies of the cosmic microwave background.
In this case, combining $B$ and $P$ does not give any appreciable benefit other than allowing a precise determination of galaxy bias.
Finally, we discuss how results depend on the binning strategy for the clustering statistics as well as on the maximum wavenumber. We also show that only considering the bispectrum monopole leads to a significant loss of information.
\end{abstract}


%%%%%%%%%%%%%%%%%%%%%%%%%%%%%%%%%%%%%%%%%%%%%%%%%%

%%%%%%%%%%%%%%%%% BODY OF PAPER %%%%%%%%%%%%%%%%%%

\section{Introduction}

The last decades have witnessed a tremendous increase in the size
of galaxy redshift catalogues that culminated in the completion
of the Two-degree Field Galaxy Redshift Survey (2dFGRS) and the Sloan Digital Sky Survey (SDSS) as well as their more recent extensions.
The scientific output of these efforts have been unprecedented and
contributed to fostering several fields of astrophysics.
The detection of baryonic acoustic oscillations in the galaxy 
two-point statistics 
\citep{Cole+2005, Eisenstein+2005}
was a major breakthrough in cosmology, as it
allowed us to measure the distance-redshift relation on large scales
and thus reconstruct the expansion history of the Universe.

Still, there is need for conducting even wider and deeper observational campaigns to address several key issues: (i) the nature of dark energy and dark matter, (ii) the neutrino masses, (iii) the statistical properties of primordial density fluctuations. These are the main science drivers of the planned next generation of surveys that will be conducted, for instance,
with the Dark Energy Spectroscopic Instrument \citep[DESI,][]{DESI1,DESI2},
the \textit{Euclid} satellite \citep{redbook} and the Square Kilometre Array  \citep[SKA,][]{SKA}. 

It is customary to extract cosmological information from galaxy catalogues
using the two-point correlation function or its Fourier transform, the power spectrum. Either of these functions fully characterize a zero-mean
Gaussian random field.
However, 
the galaxy distribution displays complex patterns characterized by
elongated filaments, compact clusters, and volume-filling underdense regions. These features are not captured by two-point statistics that do not retain information on the phases of the Fourier modes of the galaxy distribution. Therefore, if measured with sufficient accuracy
and precision, higher-order statistics like the $n$-point
correlation functions (with $n>2$) and their Fourier transforms, the polyspectra, should contain additional information.

Until recently, galaxy
redshift surveys could only provide rather noisy and imprecise measurements of
higher-order statistics \citep{Jing-Borner-1998, Frieman+Gaz-1999, Scoccimarro+2001, Verde+2002, Croton+2004, Jing-Borner-2004, Kulkarni+2007, Gaztanaga+2009, Marin2011}.
In fact, the presence or the absence of rare large-scale structures within the surveyed volume can shift the estimated statistics significantly
thus calling for the need to build
statistically representative samples that
cover larger volumes
\citep{Croton+2004, Gaztanaga+05, Nichols+06}.
For this reason, there is a lack of dedicated
tools (theoretical predictions, estimators, likelihood models) to analyse higher-order
statistics 
with respect to those specifically developed for the power spectrum.
However, the situation is gradually changing as surveys cover
unprecedentedly large volumes sampled with high galaxy number densities \citep{Gil-Marin15_fs8, Gil_Marin_fs8, Slepian+2017}.
In particular, the bispectrum will be robustly and accurately measured with the advent of the above-mentioned experiments of the next generation. 
Developing techniques for exploiting the galaxy bispectrum
is thus necessary to maximize the scientific return of these missions.

Historically, the bispectrum has been considered as a useful tool to learn about the statistical properties of the primordial density perturbations that seeded structure formation (their degree of non-Gaussianity, in particular) and to study non-linear physical processes like gravitational dynamics and galaxy biasing.  
Since these processes generate different functional dependences on the triangular configurations, they can be disentangled by fitting the measurements with theoretical templates. This procedure, for instance,  removes the degeneracy between the galaxy linear bias coefficient and the amplitude of the dark-matter perturbations invariably found in power-spectrum studies \citep[e.g.][]{Fry94, Matarrese+97,SefusattiBisp}.

Forecasts for the constraining power of the galaxy bispectrum usually determine the expected uncertainty for the bias and/or non-Gaussianity coefficients 
by assuming the main cosmological parameters are known exactly \citep{Scoccimarro+4,SefusattiKomatsu,Song15,S2,Yamauchi+2017a,Karagiannis+2018}.
This strategy has been
recently extended to modified theories of gravity \citep{Yamauchi+2017b}.
In this paper, we follow a different approach and
use the Fisher-matrix formalism to quantify 
the potential of the bispectrum as a means to extract additional cosmological information with respect to traditional power-spectrum studies. 
For surveys of the previous
generation, a similar analysis has been
presented by \citet{SefusattiBisp} who 
made forecasts for the combination of galaxy-clustering data from SDSS North with 
the analysis of the
cosmic microwave background (CMB) performed by the Wilkinson Microwave Anisotropy Probe (\textit{WMAP}).
Given the substantially improved perspectives for studies of galaxy clustering, it is imperative to update the prior investigation
by utilizing the characteristics of the forthcoming surveys. 
Recent related work focuses either on developing optimal compression algorithms for three-point statistics \citep{Byun+2017, Gualdi_2} or on detecting primordial non-Gaussianity due to the presence of massive spinning particles during inflation \citep{Morad+2018}. 
Here, we discuss the advantages (or lack thereof) of combining measurements of the galaxy power spectrum and bispectrum to constrain the standard cosmological parameters and, in particular, the dark-energy equation of state. 
In order to provide a concrete example, we focus on a \textit{Euclid}-like survey
and consider flat cosmological models dominated by dark energy and
cold dark matter (CDM) with Gaussian primordial perturbations.
We also combine the constraints from the clustering data with those from the CMB analysis by the
\textit{Planck} mission.
Apart from considering datasets of current interest,
we improve upon 
\citet{SefusattiBisp}
in multiple other ways. For instance, we (i) consider the full galaxy bispectrum in redshift space instead of its monopole moment, (ii) make forecasts for dynamical dark-energy models, and (iii) account for a more sophisticated bias expansion that also depends on the tidal field and which represents the current state of the art.
We are interested in the constraining power of two- and three-point statistics 
of the actual galaxy distribution in redshift space.
Therefore, as a first step, we neglect observational limitations that will somewhat reshuffle and degrade the information. 
For example,
we only approximately take into account
the survey geometry through our binning strategy and neglect the Alcock-Paczynski effect
\citep[as in][]{SefusattiBisp}.
These issues will be accounted for in our future work.

The paper is organized as follows. 
In Section~\ref{definitions}, we introduce our notation and define
the relevant statistical quantities.
In Section~\ref{sec:fisher}, we briefly summarize the Fisher-matrix formalism and describe the set-up of our study.
Our results are presented 
in Section~\ref{results}
and discussed in Section~\ref{disc}.
Finally, in Section~\ref{conclusions}, we conclude.

 \section{Galaxy statistics}
\label{definitions} 
\subsection{Power spectrum and bispectrum}
Given a galaxy population, we model its spatial distribution at fixed time as the discrete
sampling of a continuous random field $\rho_{\rm g}(\bx)$ which gives the local galaxy density
per unit comoving volume in the expanding Universe. We assume
that $\delta_{\rm g}(\bx)$ is statistically homogeneous, i.e. that all its connected
$n$-point correlation functions are invariant under spatial translations.
After defining the mean galaxy density $\bar{\rho}_{\rm g}=\langle \rho_{\rm g}(\bx)\rangle$ (the brackets here denote averages taken over an ideal ensemble of realisations), we introduce the dimensionless overdensity as
\be
\delta_{\rm g}( \bx) = \frac{\rho_{\rm g}(\bx)}{\bar \rho_{\rm g}}-1\;.
 \ee
We would like to decompose $\delta_{\rm g}(\bx)$ into simple oscillatory functions like plane waves. For a generic absolutely integrable function $f(\bx)$, we can write
\be 
f(\bx)=\int \tilde{f}(\bk)\, e ^{i  \bk \cdot  \bx}\,\frac{{\rm d}^3 k}{(2\pi)^3}\;,
\ee
where 
\be 
\tilde{f}(\bk)=
\int f( \bx) \, e ^{-i  \bk \cdot  \bx}\,{\mathrm{d}}^3 x
\ee
denotes the Fourier transform of $f(\bx)$.
However, $\delta_{\rm g}(\bx)$ cannot be Fourier transformed as, in almost all realisations, the integral $\int |\delta_{\rm g}(\bx)|\,{\rm d}^3x$ diverges when taken over all space. Therefore, we consider a finite region of volume $V$ and define a `sample function'
$\delta_{V}(\bx)$ such that $\delta_{V}(\bx)=\delta_{\rm g}(\bx)$ if $\bx \in V$ and
$\delta_{V}(\bx)=0$ if $\bx \notin V$. 
The power spectral density of $\delta_{\rm g}(\bx)$ can be defined as
\be
\label{eq:PV}
P(\bk)= \lim_{V\to \infty} \frac{\langle |\tilde{\delta}_V(\bk)|^2 \rangle}{V}=
\lim_{V\to \infty} \frac{\langle \tilde{\delta}_V(\bk)\,\tilde{\delta}_V(-\bk) \rangle}{V}
\;,
\ee
where the limit exists only if it is performed after taking the ensemble average.
In general, we can write 
\be 
\langle \tilde{\delta}_V(\bk)\,\tilde{\delta}_V(\bq) \rangle=\int\xi(r)\,e^{-i \bk\cdot \br}\,
{\rm d}^3r\,
\int_V e ^{-i  (\bk+\bq) \cdot  \bx}\,{\rm d}^3x\;,
\ee
where $\xi(\br)=\langle \delta_{\rm g}(\bx)\,\delta_{\rm g}(\bx+\br) \rangle$ denotes the two-point correlation function of $\delta_{\rm g}(\bx)$ and the first integral runs over all separation vectors $\br=\by-\bx$ such that 
$(\bx,\by)\in V\times V$.
Taking the limit for $V\to \infty$ and
extending the definitions above to generalized functions, we obtain
\be
\lim_{V\to \infty}
\left\langle\tilde{\delta}_{V}( \bk) \, \tilde{\delta}_{V}( \bk')\right\rangle = (2\pi)^3 \,P( \bk) \, \delta_{\rm D}( \bk +  \bk')\;,
\label{defbis}
\ee
where $\delta_{\rm D}(\bk)$ denotes the three dimensional Dirac delta distribution and
the power spectrum $P(\bk)$ is the Fourier transform of $\xi(\br)$.

Similarly, at the three-point level we can write
\be
\label{eq:BV}
B(\bk_1,\bk_2,\bk_3)= 
\lim_{V\to \infty} \frac{\langle \tilde{\delta}_V(\bk_1)\,\tilde{\delta}_V(\bk_2) 
\,\tilde{\delta}_V(-\bk_1-\bk_2) \rangle}{V}
\;,
\ee
or, equivalently,
\begin{multline}
\lim_{V\to \infty}
\left\langle\tilde{\delta}_{V}( \bk_1)\,\tilde{\delta}_{V}( \bk_2) \, \tilde{\delta}_{V}( \bk_3)\right\rangle =\\ (2\pi)^3\, B( \bk_1, \bk_2, \bk_3)\, \delta_{\rm D}( \bk_{123})\;,
 \end{multline}
 where $B( \bk_1, \bk_2, \bk_3)$ defines the galaxy bispectrum (i.e. the Fourier transform of the connected three-point correlation function)
and $ \bk_{123}=\bk_1 +  \bk_2+  \bk_3$, meaning that the bispectrum is defined only for closed triangles of wavevectors.

Different statistics (based on alternative expansions with respect to the
Fourier decompositions) need to be employed to analyse samples that cover a wide solid angle on the sky \citep[e.g.][]{Fisher+1994, Heavens-Taylor1995, Papai-Szapudi2008}.

\subsection{Redshift-space distortions}
\label{RSDs}
We infer the comoving position of a galaxy by using two observables 
(position on the sky and redshift) and by assuming that the photons
we receive from it propagate in an unperturbed Friedmann-Robertson-Walker model universe. The resulting galaxy distribution in this `redshift space' provides a distorted representation of the actual one in `real space' due to the presence of inhomogeneities and peculiar velocities.
The latter generate the largest distortions 
\citep{Jackson1972, Sargent-Turner1977, Kaiser1987, Hamilton1998} 
that dominate over other relativistic
effects \citep[see e.g.][and references therein]{LIGER} that we will
neglect in this work.

Although the galaxy distribution in real space is statistically
isotropic (implying that $P(\bk)$ only depends on the magnitude $k$ and $B(\bk_1,\bk_2,\bk_3)$ on the three values $k_1, k_2$ and $k_3$), 
redshift-space distortions (RSD) break this isotropy and introduce
some angular dependences.
In the distant-observer approximation, when galaxy separations are much smaller than the distance from the observer to the galaxies so that a single line of sight $\hat{\bs}$ can be defined for the whole sample, the power spectrum in redshift space depends on $k$ and $\mu=(\bk\cdot \hat{\bs})/k$. This result derives from the fact that density and velocity
perturbations are correlated \citep{Kaiser1987}.
Similarly, the redshift-space bispectrum depends on the line-of-sight projections $\mu_1$ and $\mu_2$ of $\bk_1$ and $\bk_2$ (as $\bk_3=-\bk_1-\bk_2$).
Therefore, the bispectrum depends on five variables, three of which
determine the shape of the triangle of wavevectors while the remaining
two indicate its orientation with respect to the line of sight.
In Appendix~\ref{App:RSDs} we discuss two different parameterizations
of the coefficients $\mu_1$ and $\mu_2$ in terms of convenient angular variables that here
we schematically denote by $0\leq\theta\leq \pi$ and $0\leq \phi< 2\pi$.

To reduce the complexity of cosmological investigations, 
the $\mu$-dependence of the galaxy power spectrum at fixed wavenumber is often expanded in
a Fourier-Legendre series \citep{Taylor-Hamilton-1996} 
\begin{equation}
P(\bk)=\sum_{\ell=0}^\infty P_\ell(k)\, {\cal L}_\ell(\mu)\;,
\label{Pmulti}
\end{equation}
where ${\cal L}_\ell(\mu)$ denotes the Legendre polynomials and the functions
\begin{equation}
P_\ell(k)=\frac{2\ell+1}{2}\,\int_{-1}^{1} P(k)\, {\cal L}_\ell(\mu)\,{\rm d}\mu
\end{equation}
are known as the `redshift-space multipoles' of the power spectrum.
In linear perturbation theory, only the monopole ($\ell=0$), quadrupole ($\ell= 2$) and hexadecapole ($\ell= 4$) do not vanish (see equation~(\ref{eq:PS}) in Section~\ref{sub:SPT} without the exponential term on the rhs).
Recent studies show that these three multipoles indeed contain the bulk of the information on the main cosmological parameters \citep[e.g.][]{Taruya+2011, Kazin+2010-12,Beutler+2014}. Therefore, a simplified inference method (with small information loss) can be engineered by only considering three functions of $k$ instead of a function of both $k$ and $\mu$.
This approach can be generalized to the galaxy bispectrum.
In fact, the dependence on the orientation of a triangle of wavevectors can be decomposed into spherical harmonics \citep{Scoccimarro1},
\begin{equation}
B(\bk_1,\bk_2,\bk_3)=\sum_{\ell=0}^\infty\sum_{m=-\ell}^{\ell} B_{\ell m}(k_1,k_2,k_3)\,Y_{\ell m}(\theta,\phi)\;,
\end{equation}
where
\begin{equation}
B_{\ell m}(k_1,k_2,k_3)=\int_{-1}^{+1}\!\!\!\int_0^{2\pi} \!\!\!\!\!\! B(\bk_1,\bk_2,\bk_3)\,Y_{\ell m}^*(\theta, \phi)\,{\rm d}\!\cos(\theta)\,{\rm d}\phi\;.
\end{equation}
A popular choice is to focus on the coefficients with $m=0$ which are often called
the `redshift-space multipoles' of the bispectrum. They satisfy
a relation similar to equation (\ref{Pmulti}) for the $\phi$-averaged bispectrum:
\begin{equation}
\int_0^{2\pi} \!\!\! B(\bk_1,\bk_2,\bk_3)\,\frac{{\rm d}\phi}{2\pi}=\sum_{\ell=0}^\infty B_{\ell 0}(k_1,k_2,k_3)\, {\cal L}_\ell(\cos \theta)\;.
\end{equation}
These multipoles are simple to estimate from a galaxy catalogue using fast Fourier transform-based methods \citep[][see also \cite{Bianchi+2015}]{Scoccimarro2} and provide a convenient procedure to compress the
bispectrum measurements into data structures of lower dimensionality. This, however, unavoidably causes loss of information. For a fixed cosmological model,
\citet{Gagrani-Samushia} show that constraints on
the velocity linear growth factor, galaxy bias coefficients and Alcock-Paczinsky parameters
based on $B_{0 0}$, $B_{2 0}$ and $B_{4 0}$ are quite similar to those derived from the
full $(\theta,\phi)$ dependence of the bispectrum. This suggests that using only the lowest-order bispectrum multipoles is not associated with a significant loss of information about
(at least) some selected cosmological parameters.
We will revisit this issue using our own results in Section~\ref{angbintest}.

For the sake of completeness, in this work, we do not compress $P(\bk)$ and $B(\bk_1,\bk_2,\bk_3)$ into their low-order multipoles and exploit their full angular dependence in redshift space. The price we pay for doing this is dealing with large
data sets and high-dimensional covariance matrices.

\subsection{Perturbative models}
\label{sub:SPT}
We model the galaxy power spectrum and the bispectrum in redshift space by
combining three ingredients: (i) Standard Perturbation Theory (SPT) 
for the growth of long-wavelength density and velocity perturbations in a single-stream collisionless fluid \citep[see][for a review]{SPT},
(ii) a galaxy bias model, and (iii) a non-perturbative phenomenological
model for RSD due to motions within virialized
structures (`finger-of-God' effect).
We only consider expressions to the lowest non-vanishing order in the perturbations.

\subsubsection{Definitions}
We consider a flat Friedmann-Lema\^itre-Robertson-Walker (FLRW) background
with expansion factor $a$ and
Hubble parameter $H$. The present-day
value of $H$ is $H_0=100\,h$ km s$^{-1}$ Mpc$^{-1}$.
We model dark energy as a barotropic fluid with equation of state
$p=w\rho c^2$ where $p$ and $\rho c^2$ denote pressure and energy density, respectively, and
$w$ is a dimensionless parameter that
can, in principle, change with $a$.

The evolution of $a$ and $H$ is regulated by Friedmann equations that can be expressed in terms of the
present-day value of the
matter density parameter
$\Omega_{\mathrm m}$ and the
dark-energy equation of state.
Neglecting the late-time contribution
from radiation, we have
\begin{equation}
\frac{H^2}{H_0^2}=\,\left(\frac{\Omega_{\mathrm m}}{a^3}+(1-\Omega_{\mathrm m})\,
\exp{\left\{-3\int_1^a \left[1 +w(x) \right]\,{\mathrm d}\ln x \right\}}
\right)\;,
\end{equation}
and the condition for the accelerated expansion of the Universe is $w<-1/3$.

On sub-horizon scales, linear density perturbations in the
matter component grow proportionally
to the growth factor $D_+$ that we 
compute by solving the ordinary differential
equation
\begin{equation}
D_+'' + \left( \frac{3}{a} + \frac{{\rm d} \ln H}{ {\rm d} \, a} \right) D_+' - \frac{3 \Omega_{\mathrm m}}{2 a^5 \,(H^2/H_0^2)}D_+ = 0\;,
\label{eq:D+}
\end{equation}
where the symbol $'$ denotes a derivative with respect to $a$.
In order to link linear density and
velocity perturbations, we introduce
the growth-of-structure parameter
\begin{equation}
f=\frac{{\rm d} \ln D_+}{{\rm d} \ln a}\;.
\end{equation}

\subsubsection{Galaxy biasing}
We adopt an Eulerian non-linear and non-local bias model to 
express the fluctuations in the galaxy density in terms of the underlying matter perturbations, $\delta(\bx)$, and the traceless tidal field with Cartesian components $s_{ij}(\bx)=(\partial_i \partial_j-\delta_{ij}\,\nabla^2/3)\,\phi(\bx)$ (where $\delta_{ij}$ denotes the
Kronecker symbol and the gravitational potential, $\phi(\bx)$, satisfies the Poisson equation $\nabla^2 \phi=\delta$).
Namely, we write
\be
\label{bias}
\delta_g(\bx)=b_1\,\delta( \bx)+\frac{b_2}{2}\left[\delta^2( \bx)-\langle \delta^2( \bx)  \rangle\right]+
\frac{b_{s^2}}{2} \left[s^2( \bx)-\langle s^2( \bx)\rangle\right]\;,
\ee
where $b_1$, $b_2$ and $b_{s^2}$ denote the linear, the non-linear and the tidal (non-local) bias parameters, respectively. 
Equation~(\ref{bias}) extends the local bias model introduced by \citet{Fry-Gaztanaga1993} to account for the anisotropy and environmental dependence of gravitational collapse \citep{Catelan+1998}.
The tidal-bias term alters the dependence of the galaxy bispectrum on the triangular configurations of the wavevectors \citep{Catelan-Porciani-Kamionkowski2000}
and a non-vanishing $b_{s^2}$ has been measured for dark-matter haloes extracted from cosmological simulations 
\citep{Baldauf+2012, Chan+12, Saito+2014, Bel+2015}.
The tidal bias is also required to ensure a proper renormalization (in the field-theory sense) of the quadratic local bias that is otherwise sensitive to short-wavelength modes of the density field that are not suitable for a perturbative analysis \citep{Mcdonald-Roy2009,Assassi+2014,Desj-Jeong-Schmidt2018}. 

Equation~(\ref{bias}) is nowadays the standard bias model for the galaxy bispectrum and is routinely used to interpret observational data
\citep{Gil-Marin15_fs8, Gil_Marin_fs8} and make forecasts for future missions
\citep{S2,Morad+2018,Karagiannis+2018}.

\subsubsection{Galaxy power spectrum and bispectrum}
We only consider expressions to the lowest non-vanishing order in the perturbations corrected with a phenomenological model for non-linear RSD. For the galaxy power spectrum in redshift space we thus write  
 \be
\label{eq:PS}
P( \bk)=Z^2_1(\bk)\,P_{\rm L}(k)\,\exp\left[-\frac{(k\,\mu\,\sigma_{\rm p})^2}{2}\right]\;,
\ee
where $P_{\rm L}$ is the power spectrum of linear matter-density fluctuations and 
\be
Z_1(\bk)=Z_1(k, \mu)= b_1+f\mu^2
\ee
accounts for linear biasing and linear RSD. The exponential term, instead,
provides a phenomenological (non-perturbative) characterization of
the suppression of power due
to non-linear velocities.
It describes virialized motions as an incoherent Gaussian scatter with (scale-independent) pairwise velocity dispersion $a\,H\,\sigma_{\rm p}$ 
(here $\sigma_{\rm p}$ is conveniently expressed in units of $\hMpc$)
and it has been shown to approximately match the results of $N$-body simulations when $\sigma_{\rm p}$ is treated as a free parameter \citep{Peacock1992, Peacock-Dodds94, BPH96}.
Note that $a\,H\,\sigma_{\rm p}$ does not coincide with the actual pairwise velocity dispersion of the galaxies 
\citep[which is scale-dependent, e.g.][]{Scoccimarro04,Kuruvilla} and
should be merely considered as a nuisance parameter of the same order
of magnitude.
It is also important to stress that, at the scales analysed in this work, the exponential term in equation (\ref{eq:PS}) is always
very close to unity and can be approximated as $1-(k\,\mu\,\sigma_{\rm p})^2/2$. 
Therefore, our results do not depend on the assumption of a Gaussian (rather than a Lorentzian)
damping factor.

Similarly, for the galaxy bispectrum we get
\begin{multline}
\label{bisp}
B(\bk_1,\bk_2,\bk_3)=
2\,\left[Z_2( \bk_1, \bk_2)\,Z_1( \bk_1) \,Z_1( \bk_2)\, P_{\rm L}(k_1)P_{\rm L}(k_2) + \rm{cycl.}\right]\\
\times\,\exp\left[-(k_1^2\mu_1^2+k_2^2\mu_2^2+k_3^2\mu_3^2)\,\frac{\sigma_{\rm p}^2}{2} \right]\;,
\end{multline}
where the cyclic permutation runs over pairs of $\bk_1, \bk_2$ and $\bk_3$ and
the second-order kernel describing the effect of non-linearities due to dynamics, biasing
and RSD is
\begin{multline}
Z_2( \bk_i, \bk_j)=\frac{b_2}{2} + b_1 F_2( \bk_i, \bk_j) + f \mu^2_{ij}G_2( \bk_i, \bk_j) \\
+\frac{f\mu_{ij} k_{ij}}{2} \left[\frac{\mu_i}{k_i} Z_1(\bk_j) + \frac{\mu_j}{k_j} Z_1(\bk_i) \right] 
+\frac{b_{s^2}}{2}S_2( \bk_i, \bk_j)\;.
\end{multline}
Here, $ \bk_{ij}= \bk_i +  \bk_j$ and $\mu_{ij}= \bk_{ij}\cdot  \hat{\bs} / k_{ij}$, while $F_2$ and $G_2$ denote the second-order kernels of the density
and the velocity fields, respectively,
    \be
 F_2( \bk_i, \bk_j)= \frac{5}{7} + \frac{m_{ij}}{2} \left( \frac{k_i}{k_j}+ \frac{k_j}{k_i}\right) +     \frac{2}{7}\,m^2_{ij}\;,
 \label{eq:F2}
 \ee
 %%%
     \be
 G_2( \bk_i, \bk_j)= \frac{3}{7} + \frac{m_{ij}}{2} \left( \frac{k_i}{k_j}+ \frac{k_j}{k_i}\right) +     \frac{4}{7}\,m^2_{ij}\;,
 \label{eq:G2}
 \ee
 where $m_{ij}=\left(  \bk_i \cdot  \bk_j \right)/\left(  k_i k_j \right)$.
Finally, the tidal kernel 
      \be
 S_2( \bk_i, \bk_j)= m_{ij}^2 - \frac{1}{3}. 
 \ee
Although equations (\ref{eq:F2}) and (\ref{eq:G2}) hold true only in an Einstein-de Sitter universe, they
provide accurate approximations in the general case \citep{Scoccimarro+98, SPT, delaBella+2017}. 
Consistently with the power-spectrum
analysis,
in equation (\ref{bisp}), we adopt a Gaussian damping function to describe 
non-perturbative contributions to RSD.
This term depends on the parameter $\sigma_{\rm p}$ that we also use for the power spectrum.
Tests conducted against $N$-body
simulations show that this is a reasonable approximation for
matter clustering
on sufficiently large scales and for redshifts $z>0.5$ \citep{Hashimoto+17}.
In this case, the best-fitting $\sigma_{\rm p}$ does not differ much from linear-theory predictions.

\subsection{Discreteness effects}
Galaxies are discrete objects and their clustering statistics are affected by shot noise.
Assuming that their distribution derives from Poisson sampling an underlying continuous density field allows us to relate the observed spectra (denoted with a tilde) with those given
in equations (\ref{eq:PS}) and (\ref{bisp}) \citep[e.g.][]{Matarrese+97}. In terms of the galaxy number density,
$\ngal$,
\be
\tilde{P}(\bk)=P(\bk)+P_{\rm shot}\;,
\label{SNP}
\ee
\begin{multline}
 \tilde{B}(\bk_1,\bk_2,\bk_3)=B(\bk_1,\bk_2,\bk_3)\\
+\left[P(\bk_1)+P(\bk_2)+P(\bk_3)\right]P'_{\rm shot}+B_{\rm shot}\;,
\label{SNB}
\end{multline}
where $P_{\rm shot}=P'_{\rm shot}=\ngal^{-1}$ and
$B_{\rm shot}=\ngal^{-2}$.

\section{Fisher matrix}
\label{sec:fisher}
\subsection{Estimators and finite-volume effects}
Actual redshift surveys cover finite comoving volumes and contain observational artefacts (gaps, masked regions, variable depth, etc.). Clustering statistics are thus measured using specifically designed estimators that minimize the impact of these features.
An estimate for $\delta_{\rm g}(\bx)$ is usually computed by weighing the contribution of each galaxy based on the selection criteria of the survey
\citep{Feldman-Kaiser-Peacock-1994}.
Schematically,
the observed galaxy overdensity can be written as $\delta_{\rm obs}(\bx)=\delta_{\rm g}(\bx)\,W(\bx)$ (where $W(\bx)$ is the window function of the survey)
so that
$\tilde{\delta}_{\rm obs}(\bk)=\int \tilde{W}(\bq)\,\tilde{\delta}_{\rm g}(\bk-\bq)\,{\rm d}^3q/(2\pi)^3$.
Therefore, an estimator for the power spectrum in redshift space can be built by replacing the ensemble average in equation (\ref{eq:PV}) with a mean taken over a finite bin of wavevectors
with similar values of $k$ and $\mu$ in a single realization:
\be
\hat{P}_i=V^{-1}\,\int_{{\cal K}_i}\tilde{\delta}_{\rm obs}(\bk)\,\tilde{\delta}_{\rm obs}(-\bk) \,\frac{{\rm d}^3 k}{K_s}\;.
\label{Pest}
\ee
Here,
$V=\int W(\bx) \,{\rm d}^3x$ denotes the effective volume of the survey and $K_s$ is the
$k$-space volume covered by the bin $\bk \in {\cal{K}}_i$.
The ensemble average of $\hat{P}_i$ is 
\be 
\langle \hat{P}_i\rangle=\int \tilde{W}(\bk_i-\bq)\,P(\bq)\,\frac{{\rm d}^3q}{(2 \pi)^3}+\mathrm{shot\ noise\ terms}\;,
\ee
and thus $\hat{P}_i$ is a biased estimator.
This reflects
the fact that plane waves (the basis functions of the Fourier expansion) are not orthonormal over a finite, non-periodic volume.
Typically, $\tilde{W}(\bk)$ shows a prominent peak at $\bk\simeq 0$ with a width of $\Delta k\sim V^{-1/3}$ (if the surveyed volume is not
elongated, otherwise $\Delta k$ coincides with the inverse of the shortest dimension).
Therefore, the power-spectrum estimator in equation~(\ref{Pest}) mixes the contributions from Fourier modes with wavenumber differences $\Delta k< V^{-1/3}$.
This is a manifestation of the uncertainty principle between conjugate variables in a Fourier transform: if the galaxy positions are confined to a region of linear size $V^{1/3}$, then the wavenumbers of the Fourier modes are `uncertain' within a range  $2\pi/V^{1/3}$.

Likewise, after introducing an estimator for the bispectrum that averages over
a set of triangular configurations ${\cal T}_i$
centred around $(\bk_1, \bk_2, -\bk_1-\bk_2)$
\be 
\hat{B}_i=V^{-1}\,\int_{{\cal T}_i} \tilde{\delta}_{\rm obs}(\bp)\,\tilde{\delta}_{\rm obs}(\bq)\,\tilde{\delta}_{\rm obs}(-\bp-\bq) \,\frac{{\rm d}^3 p\,{\rm d}^3 q}{K_\triangle}
\label{Best}
\ee
with
\be
K_\triangle=\int_{{\cal T}_i}\delta_{\rm D}(\bp+\bq+\bk)\,{\rm d}^3 p\,{\rm d}^3 q\,{\rm d}^3 k\;,
\ee
\citep{Scoccimarro2000} one finds \citep[e.g.][]{Gil-Marin15_fs8}
\begin{multline}
\langle \hat{B}_i  \rangle=\int \tilde{W}(\bk_1-\bq)\,\tilde{W}(\bk_2-\bq)\,B(\bq_1,\bq_2,-\bq_1-\bq_2)\,\frac{{\rm d}^3q_1}{(2 \pi)^3}\,\,\frac{{\rm d}^3q_2}{(2 \pi)^3}\\+\mathrm{shot\ noise\ terms}\;.
\end{multline}

Although the systematic shift of $\hat{P}_i$ and $\hat{B}_j$ due to the window function
is only noticeable on scales comparable with the extension of the survey, it needs to be accounted for in order to make unbiased inference about the cosmological parameters.
One option is to deconvolve the window function from the measured spectra
\citep{Lucy74,Baugh-Efstathiou93, Lin+96}. 
Alternatively, the theoretical models can be convolved with the window function of the survey before performing a fit to the measured spectra.
A third possibility is not to use the Fourier decomposition and expand the galaxy density in orthonormal modes that maximize the signal-to-noise (S/N) ratio given the survey geometry and the selection function (plus a fiducial model for the spectra) using the Karhunen-Lo\`eve transform \citep{Vogeley-Szalay1996, Tegmark+1997}.

For simplicity, in this work, we only approximately take into account the effects of the window function by considering $k$-bins of size $\Delta k=2\pi/V^{1/3}=k_{\rm f}$ (i.e. the expected broadening
for the primary peak\footnote{If $W(\bx)=1$ within a cube of side $L$ and 0 otherwise, then 
$\tilde{W}(\bk)=\prod_{i=1}^{3}(2/k_i)\,\sin (k_i\, L/2)$ and the main peak along each Cartesian component extends for $\Delta k=2\pi/L$ on the positive-frequency side.} 
of $\tilde{W}$ for a cubic survey volume of side $L=V^{1/3}$).
We thus compute the band-averaged power spectra and bispectra by evaluating the mean over the set of configurations that contribute to each bin.
Note that most forecast papers instead just use one characteristic configuration per bin to speed the calculation up.

\subsection{Binning strategy and covariance matrices}
\label{binningstrat}
\subsubsection{Power spectrum}
Within the distant-observer approximation,
the galaxy power spectrum in redshift space is a function of $k$ and $\mu^2$.
Therefore, we define our power-spectrum estimator using bins that run 
over a spherical shell of Fourier modes of widths $\Delta k$ and $\Delta \mu$ and central values $\bar{k}_i$ and $\bar{\mu}_i$.
In this case,
\be 
K_s=\int_{{\cal K}_i} {\rm d}^3 q= 
2\pi\,\Delta \mu\,\left[\bar{k}^2 \Delta k+\frac{(\Delta k)^3}{12} \right]\simeq
2\pi\,\Delta \mu\,\bar{k}_i^2\Delta k\;,
\ee
where the last expression on the right-hand side is valid only for narrow bins with $\Delta k \ll \bar{k}_i$.
Note that the estimator in equation~(\ref{Pest}) is symmetric between $\bk$ and $-\bk$ meaning
that, for every $\bar{k}_i$, it suffices to consider the interval $0\leq \mu \leq 1$ and partition it over the bins
of size $\Delta \mu$.

The covariance matrix of an estimator encodes information regarding
the precision to which the estimand can be measured and the correlations between estimates corresponding to different configurations.
The covariance matrix for the binned galaxy power spectrum is defined as
\begin{multline}
(C_{\rm PP})_{ij}=\langle (  \hat{P}_i -\langle \hat{P}_i\rangle)\,(  
\hat{P}_j -\langle \hat{P}_j \rangle)\rangle
=\langle \hat{P}_i\, \hat{P}_j\rangle-\langle \hat{P}_i\rangle \,\langle \hat{P}_j \rangle 
\end{multline}
and it can be decomposed in a disconnected (or Gaussian, since it is the only term present for a Gaussian random field) contribution 
and a connected (or non-Gaussian) contribution that is proportional to the trispectrum (the Fourier transform of the connected 4-point correlation function) of the galaxy distribution.
On the large scales, we are interested in, the Gaussian contribution dominates \citep{Scoccimarro-Zaldarriaga-Hui-1999, Bertolini+2016, Mohammed-Seljak-Vlah-2017} and, for narrow bins, we can write \citep{Feldman-Kaiser-Peacock-1994,Meiksin-White1999}
\be
\label{eq:Cpp} 
(C_{\rm PP})_{ij}\simeq \frac{2\,\tilde{P}_i^2}{N_P}\,\delta_{ij}\;,
\ee
where 
\be
\label{eq:Np}
N_P=\frac{K_s}{k_{\rm f}^3}\simeq
\frac{V}{(2\pi)^2}\,\bar{k}_i^2\,\Delta k\,\Delta \mu\;.
\ee
The ratio $N_P/2$ gives the number of independent fundamental Fourier cells contributing to the band averaged power spectrum. The 2 at the denominator comes from the fact that 
the density field is real valued and $\tilde{\delta}(-\bk)=\tilde{\delta}(\bk)^*$.
Note that the statistical noise of $\hat{P}_i$ reflects the survey size: larger surveys contain more independent Fourier modes that contribute to a given bin and thus are associated with smaller random errors.
Strictly speaking, equation (\ref{eq:Cpp}) is exact only for cubic volumes with periodic boundary conditions but it is reasonable to expect that, to first approximation, the covariance does not depend on the survey shape (especially for $k\gg k_{\rm f}$).
It is also worth mentioning that only the Gaussian part of $C_{\rm PP}$ is diagonal and
non-linear couplings between Fourier modes generate non-vanishing off-diagonal terms.

\begin{figure*}
	\includegraphics[width=14cm]{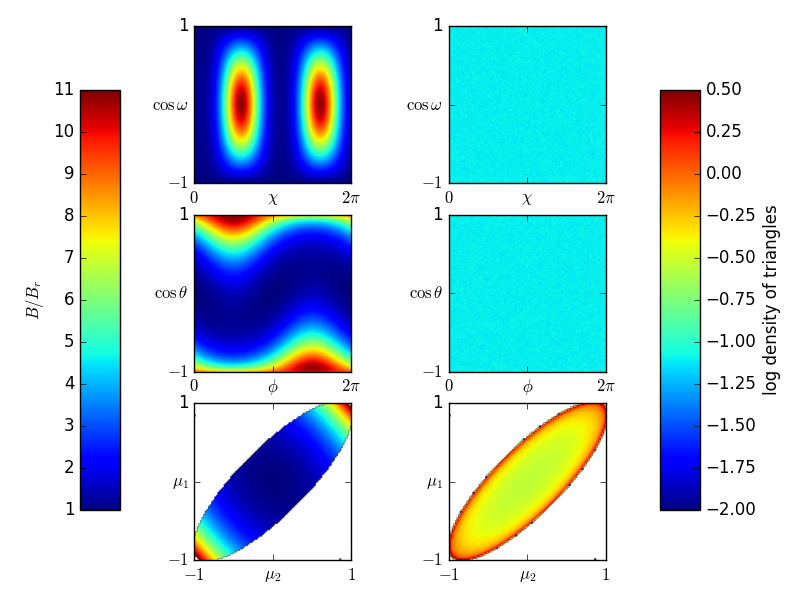}
    \caption{The panels on the left-hand side illustrate an example of how
redshift-space distortions affect the bispectrum. Shown is the ratio between the redshift-space and real-space bispectrum for a fixed triangular configuration of wavevectors
with $(k_1, k_2, k_3)=(23, 14, 10)\times 3.93\times 10^{-3} \,h$ Mpc$^{-1}$.
From top to bottom,
three different coordinate systems are used to parameterize the relative orientation of the triangle and the line of sight (see the main text and Appendix \ref{App:RSDs} for details).
The corresponding probability density
of finding a triangle with a given orientation is shown in the right-hand-side panels.
}
    \label{rsd-example}
\end{figure*}

\begin{figure}	\includegraphics[width=\columnwidth]{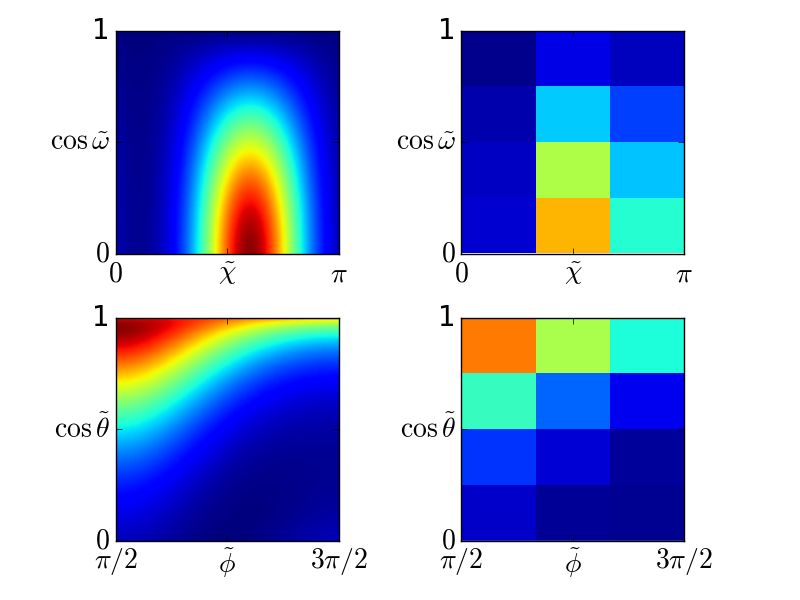}
    \caption{The redshift-space distortions displayed in Fig.~\ref{rsd-example}
    are now plotted as a function of 
the optimal angular coordinates
    $(\tilde{\omega},\tilde{\chi})$
    and $(\tilde{\theta},\tilde{\phi})$. 
We adopt infinite resolution in the left-hand-side panels and partition
parameter space into 12 
bins in the right-hand-side panels.
The colour coding is the same     
as in Fig.~\ref{rsd-example}.}
    \label{rsd-binning}
\end{figure}

\subsubsection{Bispectrum}
The galaxy bispectrum in redshift space
depends on the triangular configuration of the wavevectors and its orientation with respect to the line of sight. In this section, we show that the orientation dependence severely
complicates the analysis with respect to studies of the bispectrum in real space or the
monopole in redshift space.

We characterize the shape of a triangle using an
ordered triplet of numbers that indicate the length of its sides: $k_{\rm s}\leq k_{\rm m}\leq k_{\rm l}$. To describe its orientation, we need to use two angular variables that, for the 
moment, we denote using a generic solid angle $\Omega$.
Therefore, we define our bispectrum estimator using finite bins with central values
$\bar{k}_{\rm s}, \bar{k}_{\rm m}, \bar{k}_{\rm l}, \bar{\Omega}$
as well as widths
$\Delta k_{\rm s}= \Delta k_{\rm m}= \Delta k_{\rm l}=\Delta k$ and $\Delta \Omega$.
It follows that,
\be 
\label{Nb}
K_\triangle\simeq 8\pi^2\,\bar{k}_{\rm l}\,\bar{k}_{\rm m}\,\bar{k}_{\rm s}\,(\Delta k)^3\,
\Sigma(\bar{\Omega})\,\Delta \Omega\;,
\ee
where
$\Sigma(\bar{\Omega})\,\Delta\Omega$ denotes the fraction of triangles with fixed shape that 
populate a bin with solid angle $\Delta \Omega$, i.e. $\int_{4\pi} \Sigma(\Omega)\,{\rm d}\Omega=1$. 
Note that the right-hand side of equation~(\ref{Nb}) should be divided by 2 for degenerate triangular configurations contained in a line \citep{Mehrem09-11, ChanBlot}.

We now discuss more in detail how to parameterize the orientation of a triangle with
respect to the line of sight. To this end,
in Appendix~\ref{App:RSDs}, we introduce two different coordinate systems that we dub $(\omega, \chi)$ and $(\theta,\phi)$. 
They both define spherical coordinates but use different polar axes: 
the triangle's normal
for $(\omega, \chi)$ and one of the legs of the triangle for $(\theta, \phi)$. 
A third possibility that more closely matches
power-spectrum studies is to directly use $\mu_1$ and $\mu_2$ as indicators of the orientation of the triangle \citep[e.g.][]{Song15}.
We briefly discuss here advantages and disadvantages of these three options.
In the left column of Fig. \ref{rsd-example}, we show how RSD modify the shot-noise-subtracted galaxy
bispectrum for a fixed triangular configuration. 
From top to bottom we show 
the ratio between the redshift-space bispectrum and its real-space counterpart as a function
of $(\omega,\chi)$, $(\theta,\phi)$ and $(\mu_1,\mu_2)$.
Note that, for selected orientations,
RSD enhance the clustering signal by more than an order of magnitude.
Obviously, the size of the distortions is the same in all panels but
their overall pattern appears very different in the various coordinate systems that are
connected by non-linear transformations.

Another important quantity to analyse is the function $\Sigma(\Omega)$ that determines the noise of the bispectrum estimator as a function of the orientation of the triangles. 
By construction, the number of triangles are uniformly distributed 
in ${\mathrm d}\cos \omega\,{\mathrm d}\chi$ and ${\mathrm d}\cos \theta\,{\mathrm d}\phi$, i.e.
$\Sigma(\omega,\chi)=(4\pi)^{-1}\sin\omega$ and
$\Sigma(\theta,\phi)=(4\pi)^{-1}\sin\theta$.
On the other hand, the distribution of orientations gets more complicated when expressed in
terms of the $(\mu_1,\mu_2)$ coordinates. Using equations 
(\ref{scomu1}) and (\ref{scomu2})
to evaluate the Jacobian determinant of the coordinate transformation, we obtain\footnote{Since $\mu_2$ only depends on $\sin \phi$, there are two values of $\phi$ that give the same $\mu_2$. This explains the factor $2\pi$ in equation (\ref{eq:mu1mu2dist}).}
\begin{equation}
\Sigma(\mu_1,\mu_2)=\left(2\pi\,\sqrt{\sin^2 \xi_{12}-\mu_1^2-\mu_2^2+2\cos\xi_{12}\,\mu_1\mu_2}\right)^{-1}\;.\label{eq:mu1mu2dist}
\end{equation}
The results of a Monte Carlo simulation
obtained by randomly rotating the same triangle confirm our analytical results (the bottom right-hand panel in Fig.~\ref{rsd-example}).
Triangles only populate a finite region of the $(\mu_1,\mu_2)$ plane bounded by an ellipse
whose orientation depends on the shape of the triangles as defined by the shortest
rotation angle $\xi_{12}$ between $\bk_1$ and $\bk_2$. The density of triangles increases
considerably towards the boundaries of the ellipse. 
Regrettably, this subtlety has been missed by \citet{Song15} who, in their equation (20), assume that triangles are uniformly distributed within the entire $(\mu_1,\mu_2)$ plane. Therefore, some care should be taken when interpreting their forecasts.

In the left column of Fig.~\ref{rsd-example},
the symmetry between the triangles $(\bk_1,\bk_2,\bk_3)$ and $(-\bk_1,-\bk_2,-\bk_3)$ is evident. This corresponds to the transformations $(\omega,\chi)\to (\omega,\pi+\chi)$,
$(\theta,\phi)\to (\pi-\theta,2\pi-\phi)$ and $(\mu_1,\mu_2)\to(-\mu_1,-\mu_2)$.
In practical applications,
it makes sense, then, to select bins that combine these two configurations so that
to reduce the size of the data and, as we are about to show, also get a diagonal covariance matrix (to first approximation).
Moreover, RSD also possess an additional symmetry due to the fact that they only depend on $\sin \omega$ or
$\sin \phi$.
It is possible to `fold' the original
coordinate systems $(\omega,\chi)$
and $(\theta,\phi)$
so that to optimally exploit all
these symmetries. 
We separately discuss how to do this
in Section \ref{rsd:symmetries}
so as not to interrupt the flow of
the discussion with technicalities.
Here, it suffices to say that
we end up using two sets of variables,
$(\tilde{\omega},\tilde{\chi})$
or
$(\tilde{\theta},\tilde{\phi})$,
with the following range of variability: 
$0\leq\tilde{\omega}<\pi/2$, 
$0\leq\tilde{\chi}<\pi$,
$0\leq\tilde{\theta}<\pi/2$, and
$\pi/2\leq\tilde{\phi}<3\pi/2$.
Although they span a more compact
range,
the new coordinates
fully cover the original
parameter space shown in Fig.~\ref{rsd-example}.
The left column of Fig.~\ref{rsd-binning} illustrates how
they optimally
isolate the basic pattern that repeats four times in Fig.~\ref{rsd-example}.
It is also worth stressing that random triangular orientations are
still uniformly distributed in terms of the variables $(\cos \tilde{\omega}, \tilde{\chi})$ and
$(\cos \tilde{\theta}, \tilde{\phi})$. 
For this reason, we partition parameter space into $N_{\rm p}\times N_{\rm a}$ identical bins
of linear size $1/N_{\rm p}$ for the cosine of the polar angle (i.e. $\cos \tilde{\omega}$ or $\cos \tilde{\theta}$) and $\pi/N_{\rm a}$ for the azimuthal angle (i.e. $\tilde{\chi}$ or $\tilde{\phi}$). The right column of Fig.~\ref{rsd-binning} shows an example of how RSD
look like when $N_{\rm p}=4$ and $N_{\rm a}=3$.

The covariance matrix for the bispectrum estimator is
\begin{multline} 
(C_{\rm BB})_{ij}=\langle (  \hat{B}_i -\langle \hat{B}_i\rangle)\,
(  \hat{B}_j -\langle \hat{B}_j \rangle)
\rangle
=\langle \hat{B}_i\,\hat{B}_j \rangle-
\langle \hat{B}_i\rangle\,\langle \hat{B}_j\rangle
\end{multline}
where the indices $i$ and $j$
label bins of triangular configurations and orientations for the wavevectors.
Also in this case, the covariance can be decomposed into Gaussian and non-Gaussian contributions
that include terms up to the pentaspectrum (i.e. the Fourier transform of the connected six-point correlation function).
The Gaussian part (which is expected to dominate on large scales) receives non-vanishing contributions
whenever any one of the sides of the triangle $i$ 
is the opposite vector to any one of the sides of the triangle $j$.  Therefore, if the bispectrum bins are chosen such that
a triangle and its negative end up in the same bin, we obtain
\citep{Fry-Melott-Shandarin-1993,Scoccimarro+4,SefusattiBisp,ChanBlot}
\be
\label{eq:Cbb}
(C_{\rm BB})_{ij}\simeq 
\frac{s_{\rm B}\,V\,\tilde{P}_{i{\rm l}}\,\tilde{P}_{i{\rm m}}\, \tilde{P}_{i{\rm s}}}{N_B}\,\delta_{ij}\;, 
\ee
where the indices $(i{\rm l}, i{\rm m}, i{\rm s})$ identify the lengths and orientations of the sides of the triangular configuration $\triangle_i$ and
\be
N_B\simeq\frac{K_\triangle}{k_{\rm f}^6}\simeq\frac{V^2}{8\pi^4}\,\bar{k}_{\rm l}\,\bar{k}_{\rm m}\,\bar{k}_{\rm s}\,(\Delta k)^3\,\Sigma(\bar{\Omega})\, \Delta\Omega
\ee
gives the number of triangles falling into a bin for shapes and orientations and the coefficient $s_{\rm B}=6, 2, 1$ for equilateral, isosceles and scalene bin configurations, respectively. This number counts the matching pairs between
the sides of the bins $\triangle_i$ and $\triangle_j$. Note that the diagonal elements
of the covariance matrix are inversely proportional to the survey volume.

Finally, we consider the (rectangular) cross-covariance matrix between the estimators for the power spectrum and the bispectrum,
\begin{multline}
(C_{\rm PB})_{ij}=\langle ( \hat{P}_i -\langle \hat{P}_i\rangle ) \,
(  \hat{B}_j  -\langle\hat{B}_j\rangle )
\rangle
=\langle \hat{P}_i\,\hat{B}_j \rangle-
\langle \hat{P}_i\rangle \,\langle\hat{B}_j  \rangle
\end{multline}
which is composed of 
a disconnected part proportional to the product between $P$ and $B$ and a connected part proportional to the quadrispectrum (the Fourier transform of the connected 5-point correlation function).
\citet{SefusattiBisp} report that,
although this quantity does not have a Gaussian contribution,
it is non-negligible even on large
scales where the disconnected part dominates. In order to 
evaluate this term for our binning scheme, we need to generalize the 
expressions found in the literature that do not consider the orientation of the triangles.
A non-vanishing cross-covariance is generated by configurations in which the wavevector $\bk$
in the power-spectrum estimator, equation~(\ref{Pest}), 
coincides with (or with the reverse of) one of the legs $\bp$, $\bq$ and $-\bp-\bq$ of the triangle in the bispectrum estimator, equation~(\ref{Best}). When the bins for the power spectrum
($\bk\in {\cal K}_i$) and for the legs of the bispectrum triangles, $(\bp, \bq, -\bp-\bq)\in {\cal T}_j$,
are taken with the same criterion (for instance by only requiring that $\bar{k}_i-\Delta k/2<k<\bar{k}_i+\Delta k/2$, so that the bispectrum estimator can be labelled with three
indices $\hat{B}_{j_1 j_2 j_3}$) either zero or all triangles in ${\cal T}_{j_1 j_2 j_3}$ have, say, $\bq \in {\cal K}_i$
and the cross-covariance between $\hat{P}_i$ and $\hat{B}_{j_1 j_2 j_3}$ is given by
$(C_{\rm PB})_{ij}\simeq  2\,s_{PB}\,\hat{P}_i\,\hat{B}_j\,(\delta_{i j_1}+\delta_{i j_2}+\delta_{i j_3})/N_P$
with $s_{PB}=3,2,1$ for equilateral, isosceles and scalene triangles, respectively. 
However, to study the bispectrum in redshift space, we also bin in $\Omega$
and we need to take into account that
$\mu_{\rm m}$ and $\mu_{\rm s}$ also depend on the angular variables. 
Because of this, the $k$-space volumes spanned
by $\bk_{\rm m}$ and $\bk_{\rm s}$ within a triangular bin partially overlap with several
power-spectrum bins. Let us denote by $I_{ij_\ell}/N_B$ the fraction of triangles in ${\cal T}_j$
that have $\bk_{\ell}\in {\cal K}_i$ (i.e. a bin for $k_{\ell}$ and $\mu_{\ell}$).
 Then,
 \begin{multline}
\label{eq:CPB}
(C_{\rm PB})_{ij}\simeq 2\,s_{PB}\,
\frac{\hat{P}_i\,\hat{B}_j}{N_P\,N_B}\,\left(I_{ij_1}+I_{ij_2}+I_{ij_3} \right)\;.
 \end{multline}
Note that $\sum_i I_{ij}=N_B$, where the sum is performed over all the bins for the power
spectrum. 
For infinitesimally narrow bins, we can derive the coefficients $I_{ij}$ analytically starting from eqs. (\ref{mymu1}) and (\ref{mymu2}) or
(\ref{scomu1}) and (\ref{scomu2}).
However, 
for the broad angular bins we consider in this work, we determine them numerically.

\subsection{Survey characteristics and fiducial values}
As an example of the forthcoming next generation of galaxy redshift surveys, we consider a \textit{Euclid}-like mission. 
Within six years starting from 2021,
the \textit{Euclid} space telescope is expected to complete a wide survey that will measure
$\sim 6\times 10^7$ galaxy redshifts over $15,000$ square degrees on the sky \citep{redbook}. Low-resolution (slitless) spectroscopy in the near infrared will target the emission lines (mainly H$\alpha$) of star-forming galaxies in the approximate redshift interval $0.7< z<2.0$. 

Since only relatively small samples have been observed so far \citep[for a summary see, e.g.,][]{Pozzetti},
little is known about the population of emission-line galaxies at these redshifts. Therefore, 
we must approximate the specifics of a \textit{Euclid}-like survey by using theoretical models
that have been calibrated against the current data.
In particular, we adopt model 1 in \citet{Pozzetti} for the luminosity function of H$\alpha$-selected galaxies and assume a limiting flux of $F_{{\rm H}\alpha}>3\times 10^{-16}$ erg cm$^{-2}$ s$^{-1}$. 
In Table~\ref{tab:Euclid}, we report the corresponding galaxy number densities, $n_{\rm g}$,
as a function of redshift.
In order to facilitate comparison with
previous work, we adopt the same binning strategy as in the \textit{Euclid} Definition Study Report
\citep{redbook} and in many other forecasts for this mission \citep[e.g.][]{EuclidBP}: 14 non-overlapping redshift bins of width $\Delta z=0.1$ whose central values are linearly spaced between 0.7 and 2.0.
\begin{table}
 \caption{Specifics of a \textit{Euclid}-like survey in 14 non-overlapping redshift bins centred at $z$ and of width $\Delta z=0.1$. 
The comoving volume covered by the survey, $V$,
the galaxy number density, $n_{\rm g}$,
the characteristic halo mass, $M_0$, defined in equation (\ref{eq:hod}), and the rescaled pairwise velocity dispersion, $\sigma_{\rm p}$,
are expressed in units of 
$h^{-3}$ Gpc$^3$, $10^{-3} \, h^3$ Mpc$^{-3}$, $10^{12} h^{-1} \mathrm{M}_{\odot}$, and \hMpc, respectively.}
%%%%%%%%%%%%
\label{tab:Euclid}
 \begin{tabular}{ccccccccc}
  \hline
 $z$ & $V$ & $n_g $ & $b_1$ & $b_2$ & $b_{s^2}$ & $M_{0}$ & ${\cal N}_{\rm HO}$ & $\sigma_{\rm p}$ \\
  \hline
 0.7 & 2.82 & 2.76 &  1.18 &  -0.76 & -0.10 & 1.04 & 0.455 &4.81\\ 
 0.8 & 3.28 & 2.04 & 1.22 & -0.76 & -0.13  & 0.96 & 0.315 &4.72\\ 
 0.9 & 3.70 & 1.53 & 1.26 &  -0.75 & -0.15 & 0.88 & 0.220 &4.62 \\ 
 1.0 & 4.08 & 1.16 & 1.30&  -0.74 & -0.17 & 0.81 & 0.156 &4.51\\ 
 1.1 & 4.42 & 0.88 &  1.34 &  -0.72 & -0.19  & 0.73 & 0.108 &4.39\\ 
 1.2 & 4.72 & 0.68 &  1.38 &  -0.70 & -0.22 & 0.67 & 0.078&4.27\\ 
 1.3 & 4.98 & 0.52 & 1.42 &  -0.68 & -0.24 & 0.60 & 0.055&4.15\\ 
 1.4 & 5.20 & 0.38 &  1.46 &  -0.66 & -0.26 & 0.55 & 0.037&4.03\\ 
 1.5 & 5.38 & 0.26 & 1.50 &  -0.63 & -0.29 & 0.49 & 0.023&3.92\\ 
 1.6 & 5.54 & 0.20 & 1.54 &  -0.60 & -0.31 & 0.45 &0.017 &3.81\\ 
 1.7 & 5.67 & 0.15 & 1.58 &  -0.57 & -0.33 & 0.41 & 0.012&3.70\\ 
 1.8 & 5.77 & 0.11 & 1.62 &  -0.53 & -0.35  & 0.37 &0.008 &3.61\\ 
 1.9 & 5.85 & 0.09 &  1.66 &  -0.49 & -0.38 & 0.33 &0.006 &3.49\\ 
 2.0 & 5.92 & 0.07 &  1.70 &  -0.45 & -0.40 & 0.30 & 0.004&3.40\\
  \hline
 \end{tabular}
\end{table}

\begin{figure}
	\includegraphics[width=\columnwidth]{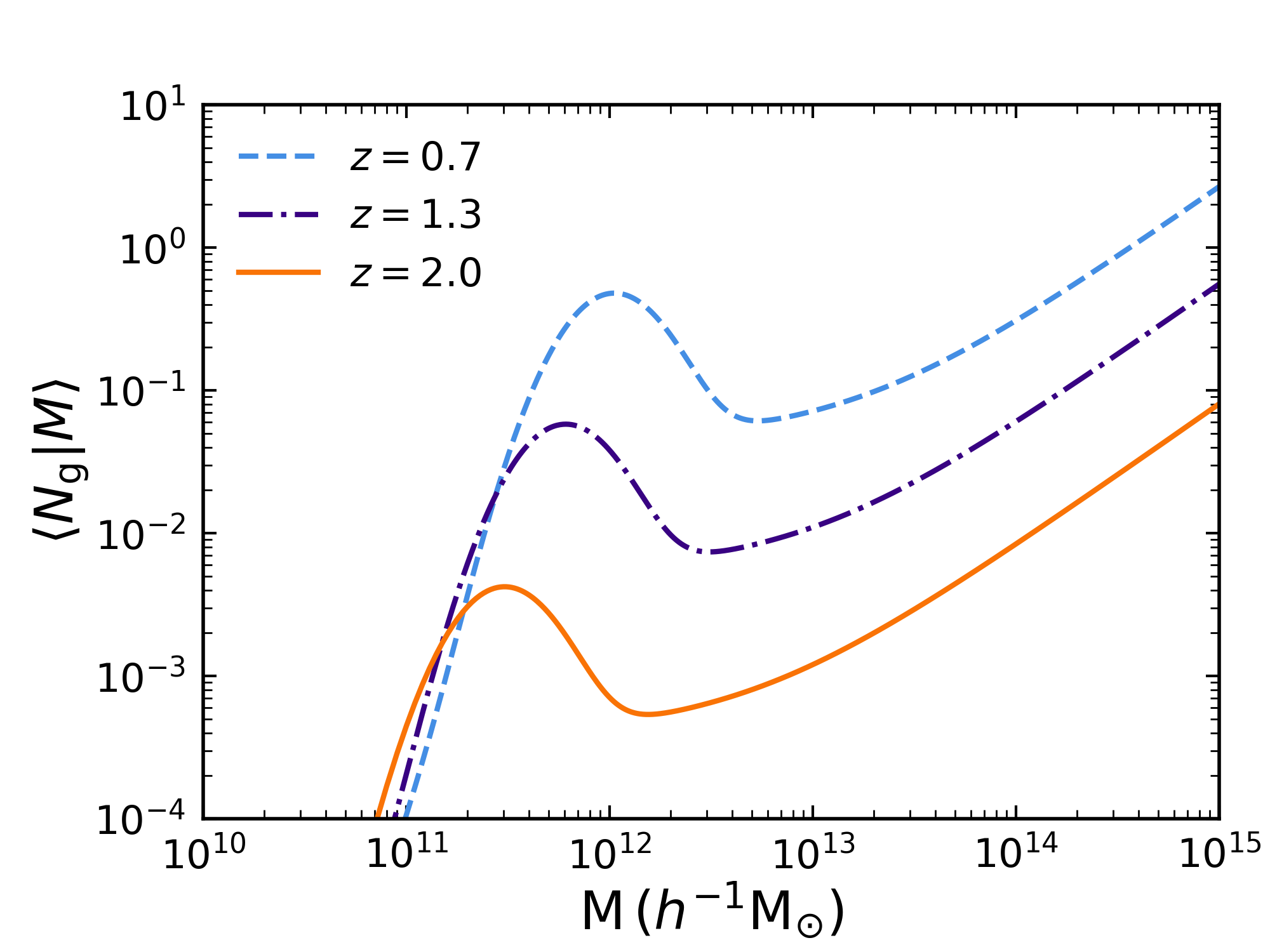}
    \caption{Halo-occupation number of the \textit{Euclid} galaxies at different redshifts.}
    \label{fig:HOD}
\end{figure}

The clustering properties of H$\alpha$ emitters at $z\sim 1$ are also very poorly constrained.
Semi-analytic models of galaxy formation combined with $N$-body simulations suggest that the linear bias parameter of the emission-line galaxies that will be detected by \textit{Euclid} should be slightly above unity at $z\sim 0.7$ and grow with redshift \citep{Orsi+2010}. An approximate fit (that we adopt) for the effective linear bias in each
redshift bin is $b_1=0.9+0.4\,z$ \citep[see appendix A in][]{Pozzetti} although observations
over two degree-sized fields at slightly higher redshifts indicate that $b_1$ could be a bit higher \citep[$b_1=2.4^{+0.1}_{-0.2}$ at $z=2.23$,][]{Geach+12}.
Determining realistic fiducial values for the quadratic and tidal bias coefficients of \textit{Euclid} galaxies requires making some additional assumptions. 
It is a basic tenet of the standard cosmological model that galaxies lie within dark-matter haloes: a central galaxy sits in the densest region of a halo while multiple satellites can be found in the outskirts.
The linear and quadratic bias coefficients of the host haloes depend on the halo mass
and redshift but
can be related to each other
by using fitting functions calibrated against $N$-body simulations, typically polynomials of second or third order \citep{bias_b2, gaztanaga}. Similarly, if halo formation is a local
process in Lagrangian space and there is no initial tidal bias, then 
\be
\label{eq:bs2}
b_{s^2}=\frac{4}{7}\,(1-b_1)
\ee
\citep{Catelan+1998,Catelan-Porciani-Kamionkowski2000,Baldauf+2012,Chan+12}.
In brief, under some reasonable assumptions, knowing $b_1$ is sufficient 
to derive $b_2$ and $b_{s^2}$ for the host haloes.
In order to extend this method to the galaxies, we model their 
halo-occupation number $\langle N_{\rm g}|M\rangle$ that gives the mean number of galaxies contained within a single dark-matter halo of mass $M$.
Uncountable studies have shown that,
for galaxies selected by luminosity in a broadband optical filter (or by stellar mass),
$\langle N_{\rm g}|M\rangle$ can be well approximated by the sum of a step function (describing central galaxies and ranging between 0 and 1) and a power law (describing satellite galaxies).
However, when galaxies are selected by the intensity of an emission line (or by star-formation rate),  $\langle N_{\rm g}|M\rangle$ is better described by a uni-modal function 
that always assumes values smaller than one (for the central galaxies) plus a power law (for the satellites).
The latter parameterization has been used 
by \citet{Geach+12} to model the observed clustering of H$\alpha$ emitters at $z\sim 2.2$
and by 
\citet{Gonzalez-Perez} to describe the population of
[\ion{O}{ii}] emitters in a semi-analytic model of galaxy formation.
We approximate their results by using a simple expression containing 
a free parameter ($M_0$) that determines the typical halo mass 
and a second one (${\cal N}_{\rm HO}\leq 0.95$) that fixes the overall normalization:
\begin{equation}
\langle N_{\rm g}| M \rangle  = {\cal N}_{\rm HO} \left(\langle N_{\rm c}| M \rangle + \langle N_{\rm s}| M \rangle\right)
\label{eq:hod}
\end{equation}
with
\begin{equation}
\langle N_{\rm c}| M \rangle =
\exp{\left\{-10\left[ \log_{10}\left(\frac{M}{M_{0}}\right)\right]^2\right\}}+0.05\, \Theta\left(\frac{M}{M_{0}}\right)\;,
\label{eq:hod1}
\end{equation}
\begin{equation}
\langle N_{\rm s}| M \rangle =
0.003\,\frac{M}{M_{0}}\, \Theta\left(\frac{M}{M_{0}}\right)\;,
\label{eq:hods}
\end{equation}
and
\begin{equation}
\Theta\left(\frac{M}{M_{0}}\right)= 1+{\rm erf}\left[2\log_{10}\left(\frac{M}{M_{0}}\right)\right]\;.
\end{equation}
The first term on the right-hand side of equation (\ref{eq:hod}) describes the halo occupation number of central galaxies while the second one refers to satellite galaxies.
Here, $M_{0}$ denotes the halo mass at which the mean number of central galaxies reaches its
maximum. 
Given the halo mass function $n(M)$ in each redshift bin \citep{SMT},
we determine $M_0$ by requiring that the effective linear bias
of the \textit{Euclid} galaxies
\begin{equation}
b_{\rm eff}= \frac{\int b_1(M)\,n(M)\,\langle N_{\rm g}| M \rangle\,{\rm d} M} 
{\int n(M)\,\langle N_{\rm g}| M \rangle\, {\rm d} M}
\label{beff}
\end{equation}
coincides with the fit given in \cite{Pozzetti}.
Using the resulting $M_0$, we then determine the effective value of $b_2$ by averaging the quadratic halo bias with weights given by the mass function and the halo occupation number
as in equation~(\ref{beff}). 
We have checked the stability of our results with respect to the parameterization of the halo mass function \citep[][and references therein]{Bhattacharya}.
Note that,
since the halo tidal bias depends linearly on $b_1$, we can obtain $b_{s^2}$ for the galaxies directly from their linear bias.
The complete set of the bias coefficients we obtain is listed in Table~\ref{tab:Euclid}.
It is worth stressing that the values of $b_2$ are always slightly less negative than (but very close to) those that would be obtained from $b_1$ by straightforwardly applying the relation between the bias parameters
that holds for dark-matter haloes. This shows that the details of the halo-occupation model
are not very important for determining $b_2$ and strengthen our confidence in the approximate
methods we have used.
For the sake of completeness, in Fig.~\ref{fig:HOD}, we plot the halo-occupation number of the \textit{Euclid} galaxies at different redshifts. The normalization
constant ${\cal N}_{\rm HO}$(unnecessary to determine the bias coefficients) is obtained by requiring that $n_g=\int n(M)\, \langle N_{\rm g}| M \rangle\, {\rm d} M$ (see Table~\ref{tab:Euclid}).

The last parameter we need to fix in order to build a fiducial model for the power
spectrum and the bispectrum of \textit{Euclid} galaxies 
is the rescaled pairwise velocity dispersion, $\sigma_{\rm p}$.
As we briefly mentioned above, 
$N$-body simulations suggest
that, at the redshifts of interest here, $\sigma_{\rm p}$ can be well approximated by linear-theory predictions \citep{Hashimoto+17}.
Therefore, neglecting velocity bias,
we write $\sigma_{\rm p}^2=2\,\sigma_\varv^2$
(where $a\,H\,\sigma_\varv$ denotes 
the 1-dimensional velocity dispersion
for the dark matter) with
\begin{equation}
\sigma^2_\varv=\frac{f^2}{3}\int_0^\infty
\frac{P_{\rm L}(k)}{k^2}\,\frac{{\mathrm d}^3k}{(2\pi)^3}=
\frac{f^2}{6 \pi^2} \int_0^\infty P_{\rm L}(k) \,{\mathrm d}k\;.
\end{equation}
Our results are summarized in Table~\ref{tab:Euclid}.

\subsection{Cosmological models}
\label{cosmo}

\begin{table*}
 \caption{Summary of the cosmological models considered in this paper. 
Here, $N_{\rm {par}}$ indicates the total number of free parameters in the fit while the last column
gives the name of the Monte Carlo generated 
Markov chains for the \textit{Planck} data we use to generate the prior. }
 \label{tab:cosm_models}
 \begin{tabular}{ccccccc}
  \hline
  Model & 
  $N_{\rm{par}}$ & Cosmology  & Bias  & Nuisance  & \textit{Planck} sample\\
  &  &  &14 $z$-bins &14 $z$-bins & \\
  \hline 
 $\Lambda$CDM &
 61 & $\Omega_{\rm{cdm}}$, $\Omega_{\rm{b}}$, $h$, $n_{\rm{s}}$, $A$ & $b_1(z)$, $b_2(z)$, $b_{s^2}(z)$& $\sigma_{\rm p}(z)$ & base-plikHM-TTTEEE-lowTEB\\ 
  $w$CDM &
  62&  $\Omega_{\rm{cdm}}$, $\Omega_{\rm{b}}$, $h$, $n_{\rm{s}}$, $A$, $w$ & $b_1(z)$, $b_2(z)$, $b_{s^2}(z)$ & $\sigma_{\rm p}(z)$ & w-base-plikHM-TTTEEE-lowTEB\\ 
  $w_0w_a$CDM &
  63& $\Omega_{\rm{cdm}}$, $\Omega_{\rm{b}}$, $h$, $n_{\rm{s}}$, $A$, $w_0$, $w_a$& $b_1(z)$, $b_2(z)$, $b_{s^2}(z)$ & $\sigma_{\rm p}(z)$ & base-w-wa-plikHM-TT-lowTEB-BAO\\
  \hline
 \end{tabular}
\end{table*}
Within the CDM scenario
with Gaussian initial conditions, we consider three classes of cosmological models characterized by different parameterizations for the equation-of-state parameter of dark energy, $w$.

We first examine
plain vanilla CDM
models with a cosmological constant,
where $w=-1$ ($\Lambda$CDM).
They are controlled by 5 parameters.
The present-day values of the density
parameters for dark matter, $\Omega_{\rm{cdm}}$, and baryons, $\Omega_{\rm{b}}$,
as well as the Hubble constant, $h$, fully determine the background. 
At the same time, we assume a power-law form for the power spectrum of 
primordial (scalar, adiabatic) curvature perturbations
\begin{equation}
{\cal P}_{\cal R}(k)=A\,\left(\frac{k}{k_*}\right)^{n_{\rm{s}}-1}\;,
\end{equation}
which is then
completely determined by the spectral index $n_{\rm{s}}$ and the 
amplitude $A$ at the pivot
scale $k_*=0.05$ Mpc$^{-1}$.

The simplest extension to $\Lambda$CDM we consider is a phenomenological model in which
$w$ stays constant with time but can assume values different from -1.
We refer to this case, where $w$ is treated as a sixth cosmological parameter, as $w$CDM.

The next level of complexity is to
use two parameters to describe
a time-varying equation of state
\citep[see e.g.][for a review]{Sahni-Starobinsky-2006}.
We adopt the popular choice of assuming
that $w$ evolves linearly with
$a$ and write 
\begin{equation}
w=w_0+w_a\,(1-a)
\label{CPL}
\end{equation}
\citep{Chevallier-Polarski2001, Linder2003}.
Here, $w_0$ gives the present-day value of the equation-of-state parameter while $w_a$ describes
its current rate of change.
Although,
these phenomenological parameters
provide a useful tool to detect
deviations from a cosmological
constant from experiments,
it is not straightforward
to map them on to physical dark-energy models \citep[e.g.][]{Scherrer2015}.
Note that equation (\ref{CPL})
describes a monotonic (and rather gentle) evolution
from the primordial value
of $w_0+w_a$ to $w_0$.

In all cases, as a fiducial model we
use the $\Lambda$CDM solution with
the best-fitting parameters for
the `TT+lowP+lensing' \textit{Planck} 2015 results \citep{Planck2015}: namely, $\Omega_{\rm{cdm}}=0.2596$, $\Omega_{\rm{b}}=0.0484$, $h=0.6781$, $n_{\rm{s}}=0.9677$, $A=2.139 \times 10^{-9}$, and $w=-1$. 
Linear transfer functions for the matter perturbations are computed using the \textsc{camb} code
\citep[https://camb.info,][]{CAMB1, CAMB2}.

\subsection{Method}
For each redshift interval, we build a data vector that combines the 
(shot-noise corrected) expectation
values for the
galaxy power spectrum and the bispectrum in the selected configuration bins.
Schematically, we write
$\mathbf{D} = (P , B)$  
and we compute the Fisher information matrix
\be
\label{eq:FM}
F_{\alpha \beta}=
\frac{ \partial {\mathbf{D}}}{\partial p_{\alpha}}
\cdot \mathbfss{C}^{-1} \cdot\frac{\partial{\mathbf{D}^T}}{\partial p_{\beta}},
\ee
where $p_\alpha$ and $p_\beta$ indicate two of the model parameters and $\mathbfss{C}$ is the block 
covariance matrix
%%%%%%%%%%%%%
\ba
\mathbfss{C} =
\left(\begin{array}{cc}
\mathbfss{C}_{\rm PP} & \mathbfss{C}_{\rm PB} \\
\mathbfss{C}_{\rm BP} &  \mathbfss{C}_{\rm BB} \\
\end{array}\right)\;,
\ea
%%%%%%%%%%%%
that can be conveniently inverted
using
%%%%%%%%%%%%%
\ba
\mathbfss{C}^{-1} = 
\left(\begin{array}{cc}
\mathbfss{C}_{\rm A} & -\mathbfss{C}_{\rm A} \mathbfss{C}_{\rm PB}\mathbfss{C}_{\rm BB}^{-1}\\
-\mathbfss{C}_{\rm BB}^{-1}\mathbfss{C}_{\rm BP} \mathbfss{C}_{\rm A}  & \mathbfss{C}_{\rm BB}^{-1}+\mathbfss{C}_{\rm BB}^{-1}\mathbfss{C}_{\rm BP}\mathbfss{C}_{\rm A} \mathbfss{C}_{\rm PB}\mathbfss{C}_{\rm BB}^{-1}\\
\end{array}\right)\;,
\ea
%%%%%%%%%%%%%%
with $\mathbfss{C}_{\rm A}=(\mathbfss{C}_{\rm PP}-\mathbfss{C}_{\rm PB}\mathbfss{C}_{\rm BB}^{-1}\mathbfss{C}_{\rm BP})^{-1}$.
We then sum the partial Fisher information matrices obtained for the different redshift intervals and invert the resulting matrix
to make a forecast for the covariance matrix of the model parameters.

As a reference case, we consider wavevectors with
$k<k_{\mathrm{max}}$ where
$k_{\mathrm{max}}=0.15\,h$ Mpc$^{-1}$.
This is for three reasons.
First, with the current state of the art, it is challenging to model non-linearities in $P$ and $B$ 
for much larger wavenumbers with an accuracy that allows applications to precision cosmology. 
\citet{Lazanu+16} have recently tested various models for the real-space bispectrum of matter perturbations
against $N$-body simulations. To the lowest non-vanishing order (tree level), SPT statistically matches the numerical results to better than 5 per cent
up to $k_{\mathrm{max}}=0.17\,h$ Mpc$^{-1}$ for $z=1$ and $k_{\mathrm{max}}=0.20\,h$ Mpc$^{-1}$ for $z=2$.
Extending the calculation to next-to-leading order (i.e. adding one-loop corrections) considerably broadens
the range of validity of the theory at $z\sim 2$.
Substantially larger values for $k_{\mathrm{max}}$ at all redshifts \citep[by up to a factor of two, see table II in][]{Lazanu+16} can also be obtained by either reorganizing the perturbative expansion \citep[e.g.][]{Matsubara2008, Crocce-Scoccimarro-Bernardeau-2012} or by adopting an effective-field-theory approach in which the influence of non-perturbative small-scale physics on to the large-scale perturbations is described with modified fluid equations whose extra parameters are calibrated against numerical simulations \citep[e.g.][]{Baumann+2012, Carrasco+2012, Angulo+2014, Baldauf+2015}.
However, accounting for galaxy biasing, RSD and discreteness effects provide additional challenges for the perturbative models and reduces their range of validity.
Secondly, the numerical inversion 
of $\mathbfss{C}$ becomes more and
more demanding with increasing 
$k_{\mathrm{max}}$. In fact,
since we use a minimal bin size of $\Delta k=k_{\mathrm f}$, we end
up dealing with very high-dimensional
matrices mainly due to the large number of possible triangle configurations for the bispectrum. 
Our default choice is to use 8 bins 
(i.e. $N_{\rm p}=4$ and $N_{\rm a}=2$) for the triangle orientations
with respect to the line of sight.
In this case, we use between
approximately 31,200 and 65,500 bispectrum bins.
Although the outcome of our study does not depend on the adopted angular coordinate system, we only show results obtained by taking bins in $\cos \tilde{\theta}$ and $\tilde{\phi}$.
A third motivation for limiting our study to $k_{\mathrm{max}}=0.15\,h$ Mpc$^{-1}$ is that non-linear effects strongly enhance the non-Gaussian contributions for all the sub-matrices that form $\mathbfss{C}$ \citep[e.g.][]{ChanBlot}. In consequence, the information content of $P$ and $B$ strongly deviates from simplistic expectations based on counting Fourier modes.
For instance, when one analyses the power spectrum, these effects lead to the so-called `translinear information plateau' \citep{RimesHamilton05,NeyrinckSzapudi07,Takahashi+2009}. Basically, with increasing $k_{\mathrm{max}}$, the cumulative information about a cosmological parameter grows until it saturates
(for $k_{\mathrm{max}}\gtrsim 0.2 \,h$ Mpc$^{-1}$). 
Only by analyzing much smaller (non-perturbative) scales ($k_{\mathrm{max}}\gg 1 \,h$ Mpc$^{-1}$) can one retrieve useful information again.
Although there are indications that the cumulative information stored in the bispectrum might saturate at smaller scales than for the power spectrum, it is also evident that, in the mildly non-linear regime, it increases at a much smaller rate than in the Gaussian approximation \citep{Kayo+2013,ChanBlot}. 
These considerations, together with the fact that the hierarchy of correlation functions (and their Fourier transforms) 
should be a rather inefficient tool to retrieve information from perturbations on fully non-linear scales \citep{Carron12,CarronNeyrinck12}, have motivated alternative approaches for retrieving the information based  on non-linear transforms and Gaussianization procedures \citep[e.g.][and references therein]{Carron+Szapudi2014}.

In Table~\ref{tab:cosm_models} we summarize
the cosmological and nuisance parameters used in our main investigation.
As detailed in Section~\ref{cosmo},
the cosmology is specified by fixing
5 to 7 variables 
depending on the adopted parameterization of the dark-energy
equation of state. 
In parallel,
for each redshift bin, we 
consider 3 bias parameters and the pairwise velocity dispersion, for a total of 56 nuisance parameters
that characterize the galaxy population under study. 
In Section~\ref{disc}, we will discuss some modifications to this set-up and their implications.

\subsection{Priors}
\label{sec:priors}
Bayesian statistics requires
adopting a prior probability distribution for the model parameters.
In this regard,
we perform our analysis in two steps.
First, we study the constraining power on cosmology 
of a \textit{Euclid}-like  survey by itself.
In this case, we use directly the
Fisher matrix to produce our forecasts. This procedure only
uses information from the likelihood
function and corresponds to 
adopting very diffuse priors 
on all the parameters.
Subsequently, we combine the results
of this first exercise
with the constraints coming from the
study of cosmic-microwave-background anisotropies performed by
the \textit{Planck} mission.
To do this, we proceed as follows.
For each of the cosmological
models introduced in Section~\ref{cosmo},
we download a Markov chain 
that samples the
posterior distribution
from the \textit{Planck} web-page\footnote{\href{https://wiki.cosmos.esa.int/planckpla2015/index.php/Cosmological_Parameters}{https://wiki.cosmos.esa.int/planckpla2015/index.php/Cosmological\_Parameters}} and compute
the corresponding covariance matrix
for the subset of cosmological parameters considered here.
We then invert the covariance matrix
and sum the result to the \textit{Euclid}-like Fisher matrix.
In practice,
we treat the \textit{Planck} results as Gaussian priors for our study
of galaxy clustering.
The exact names of the files we use
are reported in Table~\ref{tab:cosm_models}. Note that, for the $w_0w_a$CDM models,
we use a combination of current CMB 
and galaxy-clustering data.

\section{Results}
\label{results}

\subsection{Signal-to-noise ratio}
\begin{figure}
	\includegraphics[width=\columnwidth]{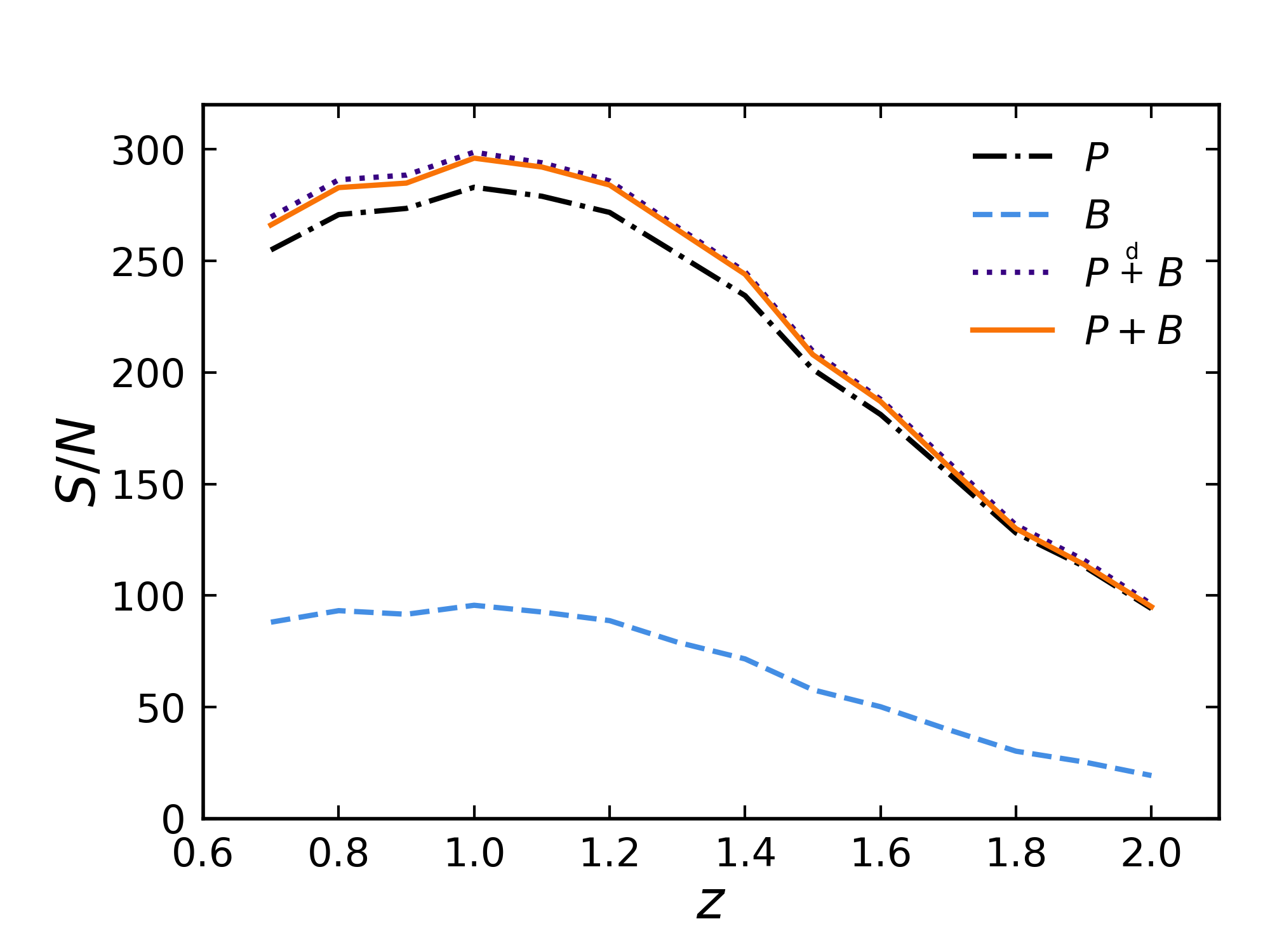}
    \caption{Signal-to-noise ratio for measurements of the galaxy power spectrum and the bispectrum in a 
    {\textit{Euclid}}-like survey as a function of redshift.
We show results for the redshift-space power spectrum (dot-dashed), the redshift-space bispectrum (dashed) and their combination (solid). For comparison, we also
display the S/N computed by neglecting the cross-covariance between $P$ and $B$ (dotted).
}
    \label{fig:stn}
\end{figure}
%%%%%%%%%%%%%%%%%%%%%%%%%%%%%%%%%%%%%%%%%%%%%%%%%
\begin{figure*}
	\includegraphics[width=14cm]
{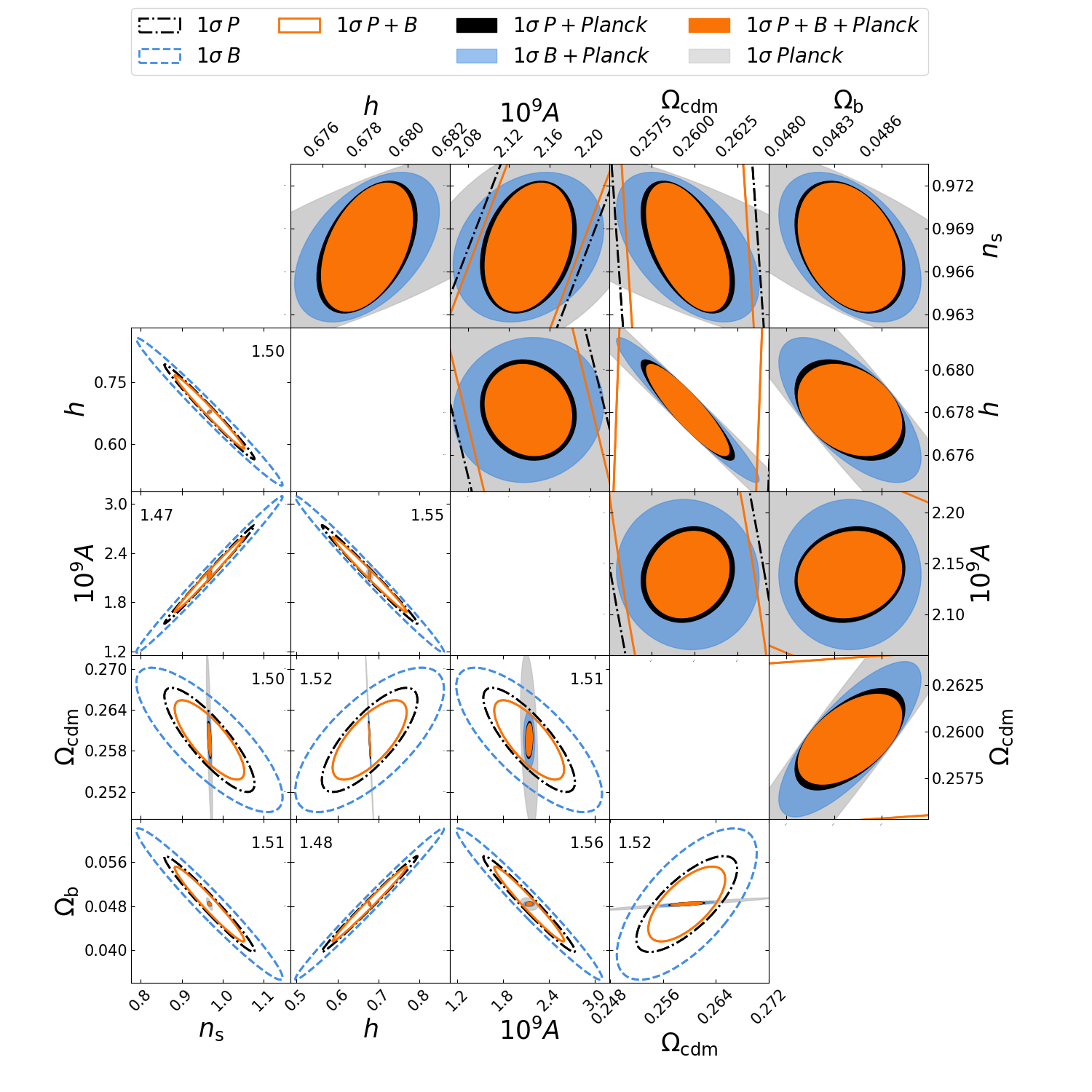}	
    \caption{
Joint 68.3 per cent
credible regions for all pairs
of cosmological parameters of the
$\Lambda$CDM model.
Different linestyles indicate
the forecast for a \textit{Euclid}-like survey based on different
observables: namely,
the power spectrum (dot-dashed), the bispectrum (dashed), and their combination (solid). 
The numbers indicate the ratio between the areas enclosed within the dot-dashed and the solid lines.
The shaded areas highlight the
credible regions obtained by also considering the \textit{Planck} priors introduced in Section~\ref{sec:priors}.
The colour coding is indicated by the top labels. The panels below the diagonal offer a panoramic view
while those above the diagonal zoom in for a close up of the central regions.}
    \label{fig:el_cdm}
\end{figure*}
%%%%%%%%%%%%%%%%%%%%%%%%%%

\begin{figure*}
	\includegraphics[width=14cm]{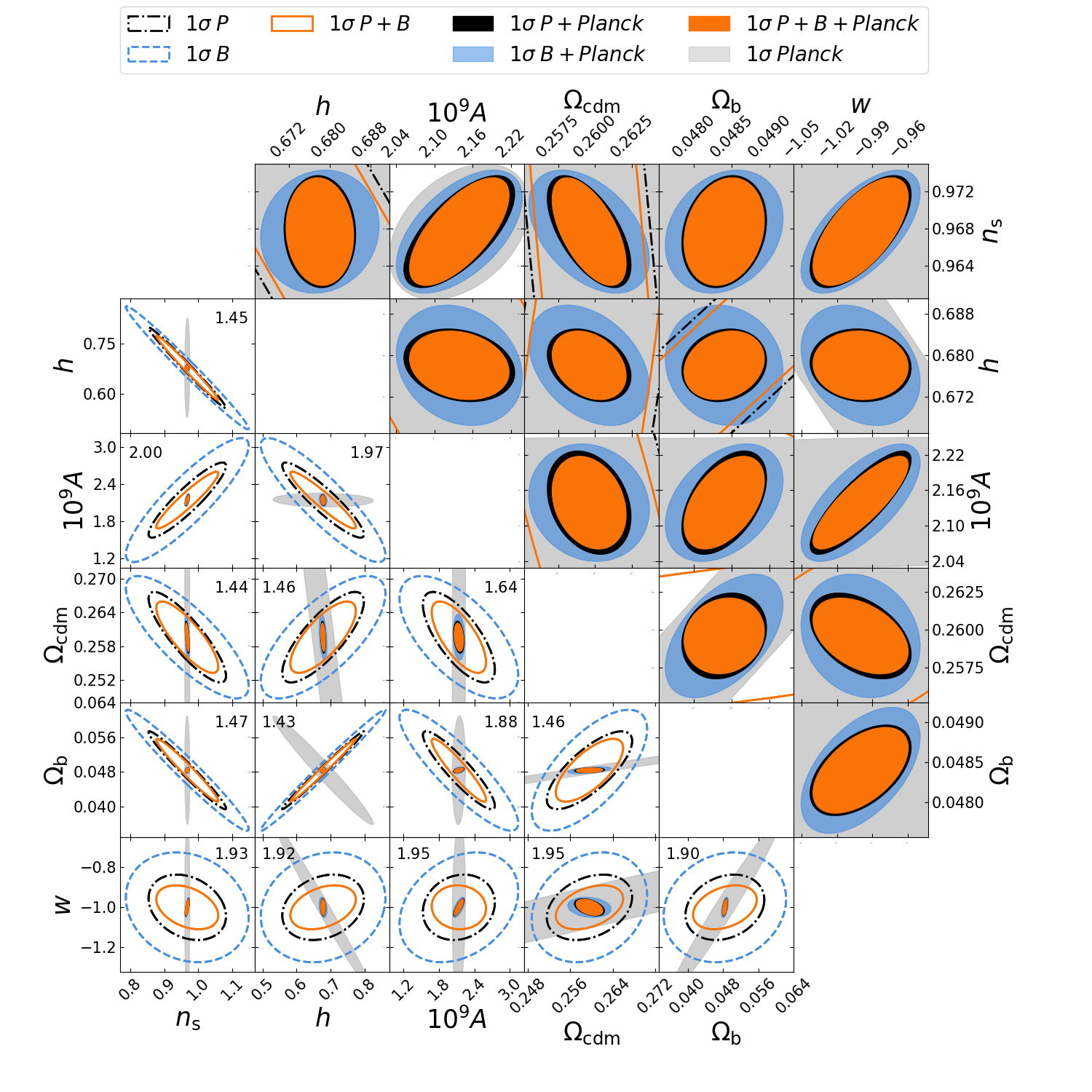}
    \caption{As in Fig.~\ref{fig:el_cdm} but
    for the $w$CDM model.}
   \label{fig:el_wcdm}
\end{figure*}
%%%%%%%%%%%%%%%%%%%%%%%%%%%%%%

\begin{figure*}
\includegraphics[width=14cm]
{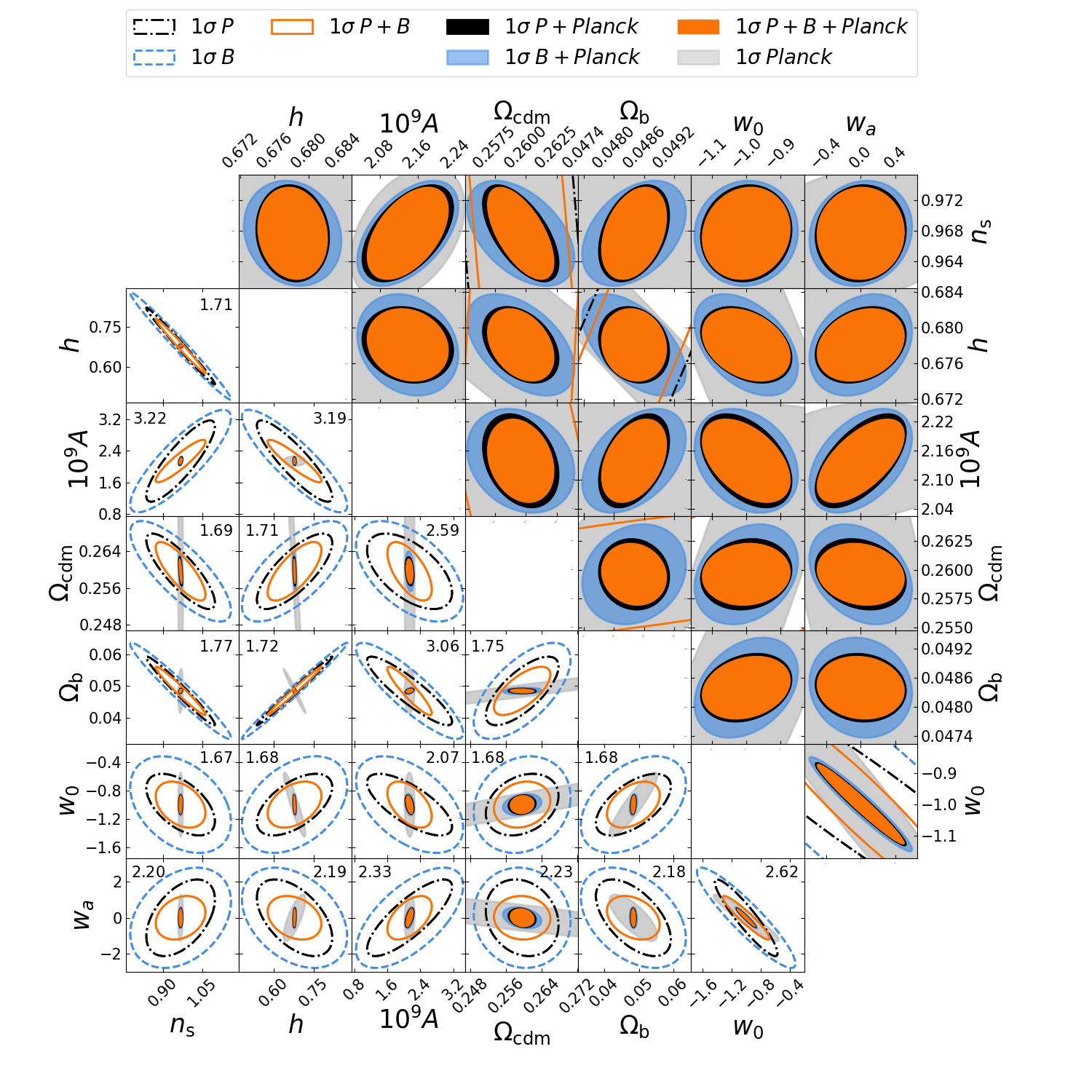}
    \caption{As in Fig.~\ref{fig:el_cdm} but
    for the $w_0w_a$CDM model.}
    \label{fig:el_w0wacdm}
\end{figure*}

In Fig.~\ref{fig:stn}, we quantify the statistical significance with which the redshift-space power spectrum and bispectrum of \textit{Euclid} galaxies will be measured. We plot
the S/N ratio
\be
\left(\frac{S}{N}\right)^2 = {\mathbf{D}}\cdot{\mathbfss C}^{-1} \cdot{\mathbf{D}}^T\;,
\ee
as a function of redshift (solid).
We also show individual results for $P$ (dot-dashed) and for $B$ (dashed) as well as
for their combination when  the cross-covariance ${\mathbfss C}_{PB}$ is assumed
to vanish (dotted).
Thanks to the huge volume covered by the \textit{Euclid}-like survey,
both the power spectrum and the bispectrum are clearly distinguishable from noise with
high confidence (note that this is not true for the single triangular configurations of the bispectrum whose measurement, given our narrow bins in $k$, is almost always dominated by noise). The global $S/N$, however, rapidly drops for $z>1.2$ 
mainly due to the decreasing galaxy number density. In spite of the very large number
of triangular configurations, we consider that the S/N ratio for $B$ is always a factor of 2.5-3 times smaller than for $P$. 
Finally, we note that neglecting the cross-covariance between $P$ and $B$, as in some previous studies \citep[e.g.][]{Karagiannis+2018}, only slightly overestimates the total $S/N$ at the lowest redshifts \citep[see also][]{Song15}. This is a consequence of the fact that we only consider quasi-linear scales where $P$ and $B$ are weakly correlated. The differences become more marked if the analysis is extended 
to smaller scales \citep{Byun+2017,ChanBlot}.

\subsection{Cosmological parameters}

In Figs.~\ref{fig:el_cdm}, 
\ref{fig:el_wcdm}, and \ref{fig:el_w0wacdm}, we show the
results of our forecasts for the 
$\Lambda$CDM, $w$CDM and
$w_0w_a$CDM models, respectively.
Shown are the joint $68.3$ per cent 
credible regions
for all possible pairs of cosmological parameters
obtained after
marginalizing over all the remaining
model parameters.
The bottom-left area of the figures
is tailored to display the likelihood
contours obtained
from a {\textit{Euclid}}-like survey.
Dot-dashed, dashed, and solid lines
show the constraints coming from 
the galaxy power spectrum, the bispectrum, and their combination, respectively.
In each panel, we report the ratio between the areas enclosed within the dot-dashed and the solid curves. These numbers show that the benefit of combining two- and three-point statistics
becomes more marked for the models that include a larger number of free parameters as there
are more degeneracies to break.
On the other hand, the narrow-shaded 
regions
highlight the credible regions
obtained by also considering the {\textit{Planck}} priors introduced in Section~\ref{sec:priors}.
The top-right areas of the figures zoom in to display the combined results more clearly.

The corresponding marginalized errors
for each single variable 
are reported in Table~\ref{tab:cdm2}.
In general, the bispectrum provides
similar, but slightly worse, constraints than the power spectrum.
Also, the orientation of the 
likelihood contours is very similar
between the two probes. Therefore,
the combination of these two- and three-point statistics leads to a non-negligible but moderate gain in the determination of the cosmological parameters.

Adding the {\textit{Planck}} prior
breaks degeneracies in the models
by imposing strong constraints on $n_s, A,$ as well as on various combinations of $\Omega_{\rm b}, h$. In consequence, the
parameters that describe the dark-energy equation of state are determined much more precisely.
Once combined with {\textit{Planck}},
the galaxy power spectrum and the
bispectrum give very similar constraints on the cosmological parameters. In this case, combining two- and three-point statistics provides only minimal advantages for the cosmology sector but yields a precise measurement of galaxy bias (see Section~\ref{gbp}).

Table~\ref{tab:cdm2} also
shows that the forecast obtained
by neglecting the cross-covariance between $P$ and $B$ is only slightly optimistic with respect to the full analysis. 
This result validates previous studies that do not consider $\mathbfss{C}_{\rm BP}$ (provided that they focus on sufficiently large scales).
Note that the numerically challenging inversion of the covariance matrix in equation~(\ref{eq:FM}) becomes
trivial when $\mathbfss{C}_{\rm BP}=0$.

\begin{table*}
 \caption{
Expected marginalized $1\sigma$ errors
(i.e. half of the 68.3 per cent credible-interval size) 
for the cosmological parameters 
in the $\Lambda$CDM, $w$CDM, and $w_0w_a$CDM models obtained considering a \textit{Euclid}-like  survey
(left) and its combination with \textit{Planck} 
priors (right).
The different columns display results obtained from the galaxy power spectrum, $P$,
the bispectrum, $B$, and their combination, $P+B$.
We also show forecasts computed by
neglecting the cross-covariance $\mathbfss{C}_{\rm BP}$ that we indicate with the symbol $P \overset{\text{\rm{d}}}{+} B$.
Note that, to ease the presentation of the results, the parameters have been rescaled by a multiplicative factor as indicated in the leftmost column of each sector. The bottom row gives the figure of merit for the dark-energy parameters $w_0$ and $w_a$.
}
 \label{tab:cdm2}
 \begin{tabular}{lcccclcccc}
   \hline
\multicolumn{5}{|c|}{ \textit{Euclid}-like alone} & \multicolumn{5}{|c|}{\textit{Euclid}-like with \textit{Planck} prior}  \\
 & $P$ &  $B$ &  $P \overset{\text{\rm{d}}}{+} B$ & $P+B$ & & $P$ &  $B$ &  $P \overset{\text{\rm{d}}}{+} B $ & $P+B$\\
\hline
\multicolumn{10}{|c|}{$\Lambda$CDM}  \\
  \hline 
   $10 \, n_{\rm{s}} $  &      0.72 &      1.17 &      0.53 &      0.56 & $10^3 \, n_{\rm{s}}  $ &      3.02 &      3.43 &      2.91 &      2.93 \\ 
$10 \, h  $  &      0.76 &      1.19 &      0.56 &      0.59 &   $10^3 \, h  $  &    1.55 &      2.24 &      1.38 &      1.41 \\ 
 $10^{10} A  $ &      3.95 &      6.28 &      2.87 &      3.01 &   $10^{11}  A  $ &    3.09 &      4.86 &      2.60 &      2.80 \\ 
 $ 10^3\, \Omega_{\rm{cdm}}  $  &      5.01 &      6.96 &      3.56 &      3.82 &  $10^3 \, \Omega_{\rm{cdm}}  $  &     1.79 &      2.73 &      1.54 &      1.58 \\ 
$10^3 \, \Omega_{\rm{b}}  $ &      5.71 &      9.02 &      4.26 &      4.46 &   $10^4 \, \Omega_{\rm{b}}  $ &    2.27 &      2.94 &      2.12 &      2.14 \\ 
  \hline
  \multicolumn{10}{|c|}{$w$CDM}  \\
  \hline
  $10 \, n_{\rm{s}}  $  &      0.75 &      1.19 &      0.56 &      0.60 & $10^3 \, n_{\rm{s}}  $ &      3.96 &      4.35 &      3.78 &      3.81 \\ 
$10 \, h  $    &      0.79 &      1.21 &      0.59 &      0.63 &   $10^3 \, h  $  &    4.61 &      7.64 &      4.27 &      4.35 \\ 
 $ 10^{10}  A  $ &      3.99 &      6.58 &      2.87 &      3.01 &   $10^{11} A  $ &    5.73 &      6.52 &      5.08 &      5.14 \\ 
 $10^3 \, \Omega_{\rm{cdm}}  $  &      5.28 &      7.14 &      3.83 &      4.17 & $10^3 \, \Omega_{\rm{cdm}}  $  &      1.86 &      2.68 &      1.61 &      1.64 \\ 
$10^3 \, \Omega_{\rm{b}}  $ &      5.87 &      9.13 &      4.45 &      4.74 &  $10^4 \, \Omega_{\rm{b}}  $ &     3.69 &      5.13 &      3.45 &      3.49\\ 
  $10 \, w  $ &      1.07 &      1.80 &      0.71 &      0.72 &   $10^2 \, w  $ &      2.80 &      3.30 &      2.61 &      2.64 \\ 
 \hline
   \multicolumn{10}{|c|}{$w_0w_a$CDM}  \\
  \hline
  $10 \, n_{\rm{s}}  $ &      0.86 &      1.26 &      0.58 &      0.62  & $10^3 \, n_{\rm{s}}  $ &      4.13 &      4.52 &      3.92 &      3.96 \\ 
$10 \, h  $  &      0.93 &      1.30 &      0.62 &      0.66 &   $10^3 \, h  $  &    2.78 &      3.74 &      2.65 &      2.67 \\ 
 $10^{10}  A  $ &      6.78 &      8.55 &      3.48 &      3.54 &  $10^{11}  A  $ &     6.32 &      7.07 &      5.62 &      5.67 \\ 
 $10^3 \, \Omega_{\rm{cdm}}  $  &      5.37 &      7.16 &      3.83 &      4.17 & $10^3 \, \Omega_{\rm{cdm}}  $  &      2.00 &      2.85 &      1.71 &      1.77 \\ 
$10^3 \, \Omega_{\rm{b}}  $ &      7.12 &      9.99 &      4.75 &      5.02 &   $10^4 \, \Omega_{\rm{b}}  $ &    4.59 &      6.81 &      4.21 &      4.26 \\ 
  $10 \, w_0  $ &      2.85 &      4.47 &      2.00 &      2.13 &   $10^2 \, w_0  $ &    8.61 &      9.88 &      8.28 &      8.38 \\ 
  $ \, w_a$ &      1.40 &      1.83 &      0.78 &      0.79 &    $10 \, w_a  $ &    3.40 &      3.88 &      3.26 &      3.29 \\ 
  \hline
      FoM$(w_0w_a)$ & 6.66 & 3.03 & 18.10 & 17.43 & FoM$(w_0w_a)$ & 147.06 & 93.32 &  166.71 & 162.49\\ 
\hline
 \end{tabular}
\end{table*}

\subsection{Galaxy bias}
\label{gbp}
\begin{figure*} 
	\includegraphics[width=\columnwidth]{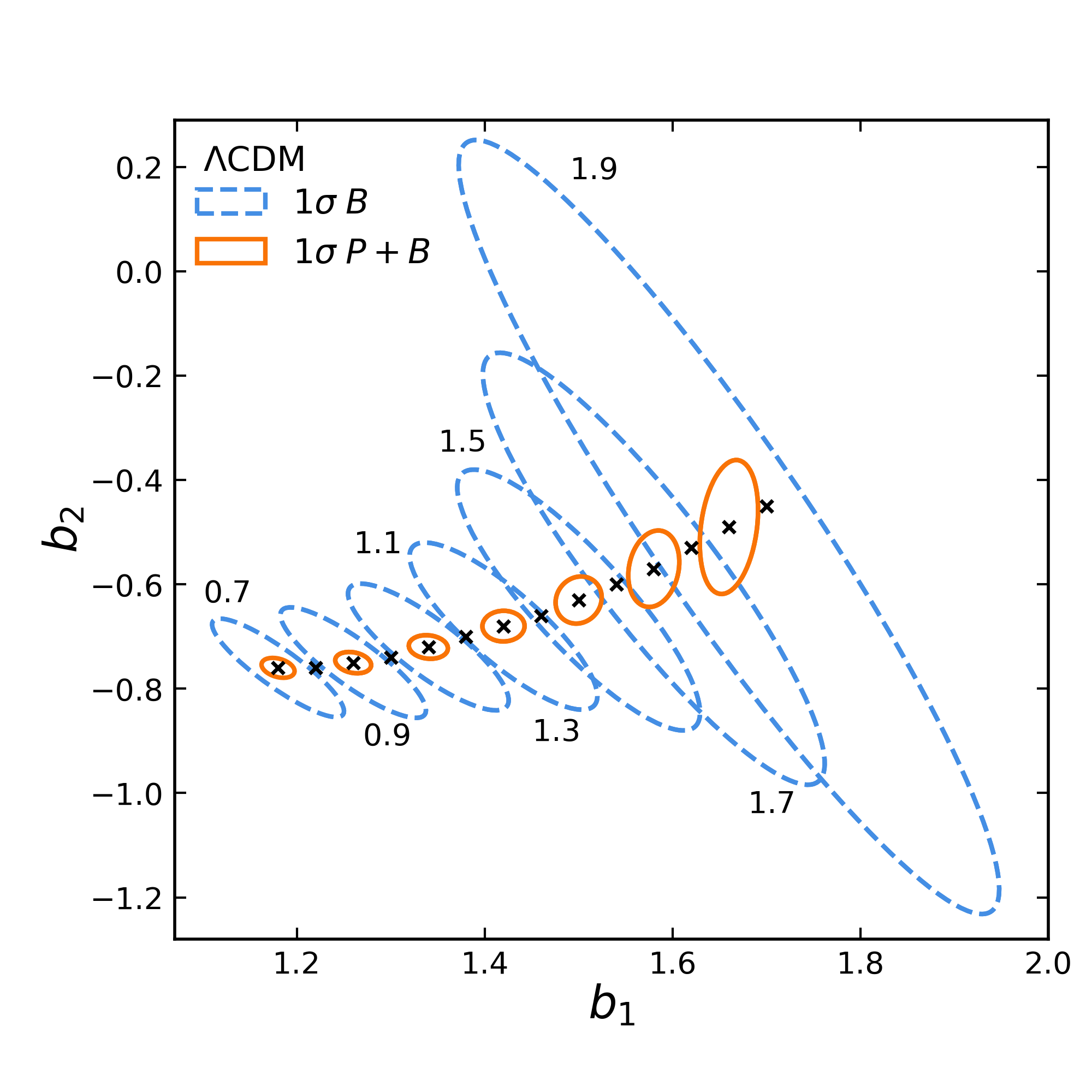}
\includegraphics[width=\columnwidth]
{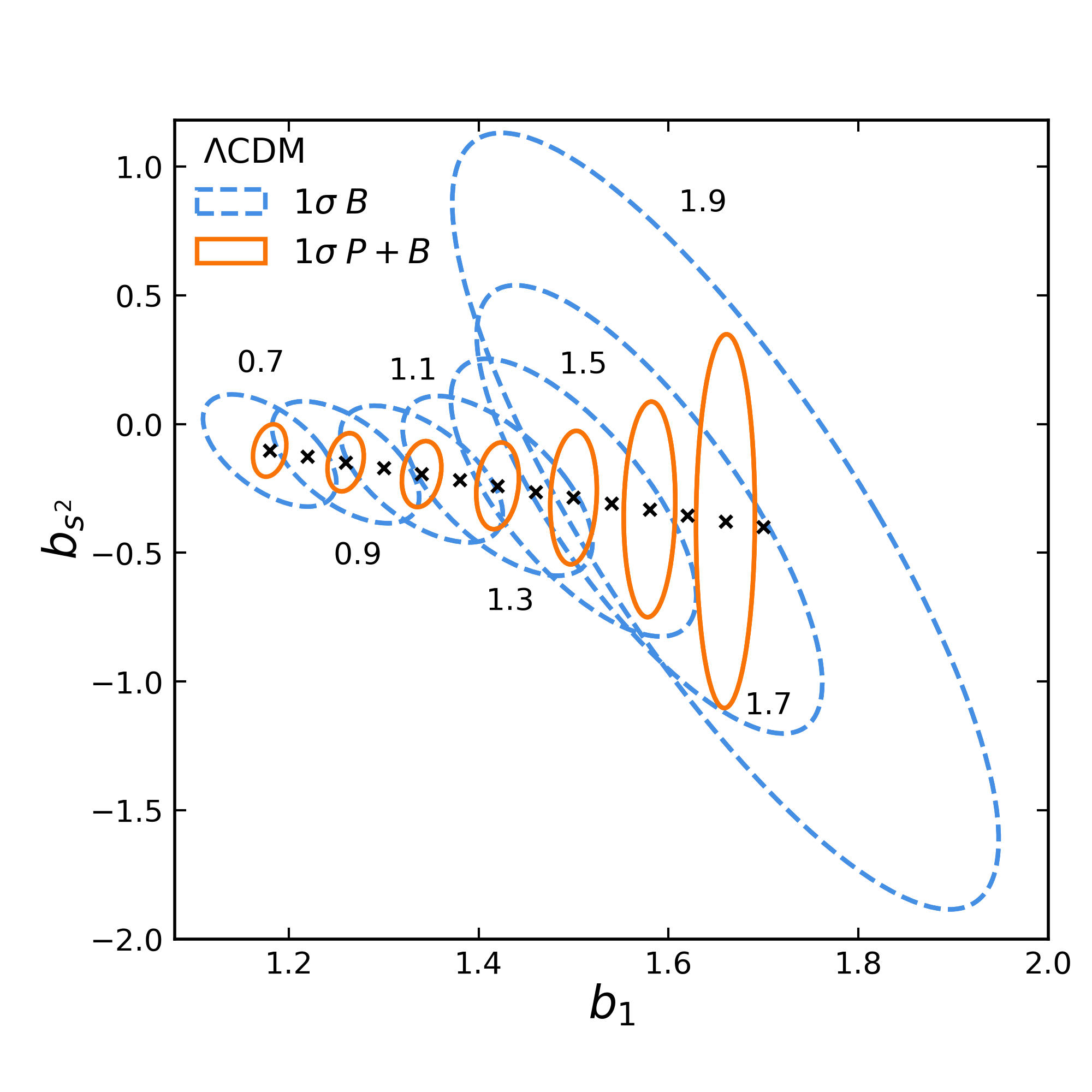}
  \caption{Joint 68.3 per cent
  credible regions for pairs
  of bias parameters determined
  using the bispectrum (dashed) and
  the combination between the power spectrum and the bispectrum
  (solid) for a \textit{Euclid}-like survey. 
To improve readability, we mark with crosses the fiducial values for all redshift bins but we show the credible regions only for alternate bins.   
  The mean redshift for the sample increases from left to right.
The numerical labels indicate the central value of each redshift bin and are located in proximity of the corresponding contours to help identify them.} 
  \label{fig:el_b1b2} 
\end{figure*}

\begin{figure} 
  	\includegraphics[width=9cm]{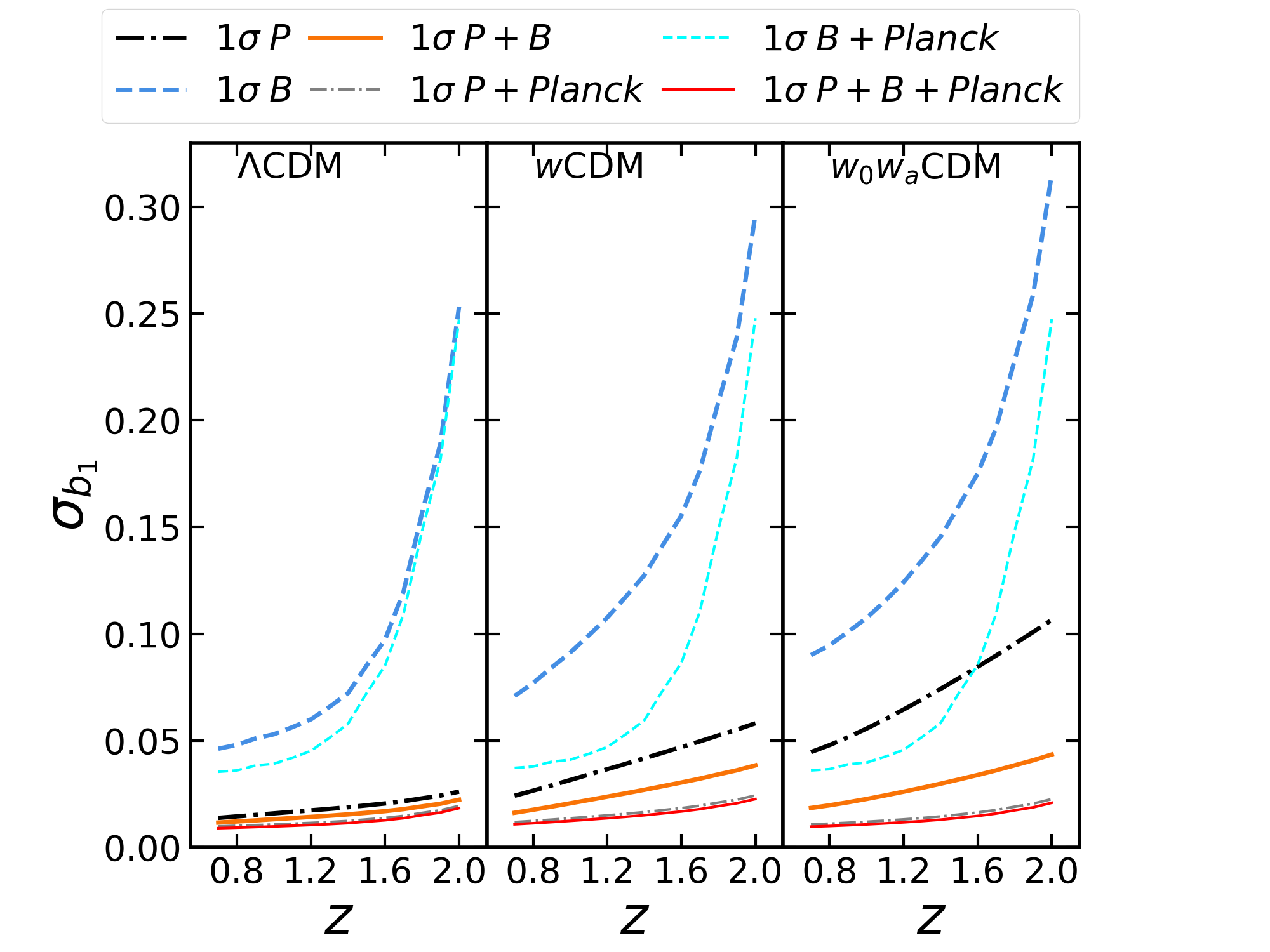}
  \caption{Forecast $1\sigma$ errors
  for the linear bias parameter as
  a function of redshift. Shown are the results for a \textit{Euclid}-like survey (with and without \textit{Planck} priors) based on 
   the galaxy power spectrum (dot-dashed), the bispectrum (dashed), and their combination (solid). Line style and thickness are indicated by the top labels.} 
  \label{fig:b1} 
\end{figure}

%%%%%%%%%%%%%%%%%%%%
Being able to accurately measure 
non-linear galaxy bias is considered one of the classic advantages of bispectrum studies.
In Fig.~\ref{fig:el_b1b2}, we present forecasts for the uncertainty with
which an \textit{Euclid}-like survey
can determine the bias parameters
in a $\Lambda$CDM model (results are similar for the other cases considered in this paper).
The bispectrum provides tight constraints on the bias coefficients
at low redshift but it does not
contain enough information to 
uniquely determine them at higher redshifts where estimates of
$b_1$ and $b_2$ (and, to a lesser degree, $b_1$ and $b_{s^2}$) are
degenerate.
Simultaneously fitting the power spectrum and the bispectrum strongly improves the situation. In fact,
the power spectrum more tightly constraints $b_1$ (see also Fig.~\ref{fig:b1}) and this is enough
to break the degeneracies with $b_2$ and $b_{s^2}$.
Combining the two probes,
leads to even smaller $b_1$ errors, especially for the $w_0w_a$CDM model
(rightmost panel in Fig.~\ref{fig:b1}).
It is worth stressing that,
in a power-spectrum study,
the error on $b_1$ 
correlates with that on most cosmological parameters 
while cosmology-bias cross-correlations are weaker
for the bispectrum.

Fig.~\ref{fig:el_b1b2} shows that
the combination of power spectrum
and bispectrum should provide
rather tight constraints in the
$(b_1, b_2, b_{s^2})$ space 
that could be used to derive the halo occupation properties of
the galaxies.
In fact, empirically measuring 
deterministic
relations between $b_1$ and $b_2$
as well as between $b_1$ and $b_{s^2}$ would 
shed light on the nature of the biasing process. For instance, measuring a negative
$b_{s^2}$ at all redshifts in accordance with equation~(\ref{eq:bs2})
would provide evidence in favour
of a local biasing process in
Lagrangian space.

%%%%%%%%%%%%%%%%%
%%%%%%%%%%%%%%%%%%%%%%%%%%%%%%%%%%%%%%%%%%%%%%%%%%%%%%%%%%%%%%%%%%%%%%%%%%
%%%%%%%%%%%%%%%%%%%%%%%%%%%%%%%%%%%%

\subsection{Figure of merit for dark-energy constraints}
Since the report of the dark-energy
task force \citep[DETF,][]{Albrecht+},
it is customary
to compare cosmological probes
in terms of a conveniently defined figure of merit (FoM), i.e. a single number 
summarizing the strength of the constraints that can be set on to
the model parameters that describe
dynamic dark energy.
For the $w_0w_a$CDM model, we adopt
the definition \citep{FoM, Mortonson+2010}
\be
\label{eq:FoM}
\textrm{FoM}=\frac{1}{ \sqrt{\textrm{det}\, \textrm{Cov}(w_0,w_a)}},
\ee
where $\textrm{Cov}(w_0,w_a)$ denotes the $2\times 2$ covariance matrix
for the errors on $w_0$ and $w_a$
(note that our definition 
is a factor of $6.17\pi$ larger than the DETF FoM that is defined as the reciprocal of the area in the $w_0$-$w_a$ plane that encloses the 95 per cent credible region).
Our results are reported in the last
row of Table~\ref{tab:cdm2}.
We find that the galaxy power spectrum in a \textit{Euclid}-like 
survey gives an FoM that is more
than two times larger than for 
the bispectrum. However, combining two- and three-point statistics improves the FoM by a factor of $2.6$ with respect to considering the power spectrum only\footnote{The corresponding factors for other combinations of cosmological
parameters can be directly read in the bottom left-hand panels of Figs. \ref{fig:el_cdm}, \ref{fig:el_wcdm}, and \ref{fig:el_w0wacdm}.}.
This promising result is, however, weakened by considering
the current CMB+clustering constraints as a prior. In this case, adding the bispectrum only improves the FoM by 11 per cent.
The reason for this behaviour is as follows. The improvement for the \textit{Euclid} data
mainly derives from partially breaking the degeneracy between 
$b_1$ and the amplitude of $P$ and $B$ for all redshift bins.
As we have shown in the previous Section, combining $P$ and $B$ allows a much better determination of the linear bias parameters at all redshifts
(the marginalized errors shrink by a factor between 2 and 3). These 14 parameters are degenerate with the amplitudes of the clustering signals that depend on both $A$ and the linear growth factors (thus on $w_0$ and $w_a$)\footnote{We have checked that, if the linear bias coefficients are kept fixed at their fiducial value, the FoM for the dark-energy parameters only improves by a factor of 1.25 when $P$ and $B$ are combined.}.
Once the \textit{Planck}'s data are taken into consideration, $A$ is extremely well determined and the constraints on $b_1, w_0$ and $w_a$ do not improve significantly by adding the galaxy bispectrum to the power spectrum.

\section{Discussion}
\label{disc}

In this section, we study how 
modifications to our standard setup
influence the forecast results. 
For simplicity, we only consider the
$\Lambda$CDM model and focus on the redshift bin centred at $z=1$.

\subsection{Dependence on the bin width $\Delta k$}
\begin{figure} 
	\includegraphics[width=\columnwidth]{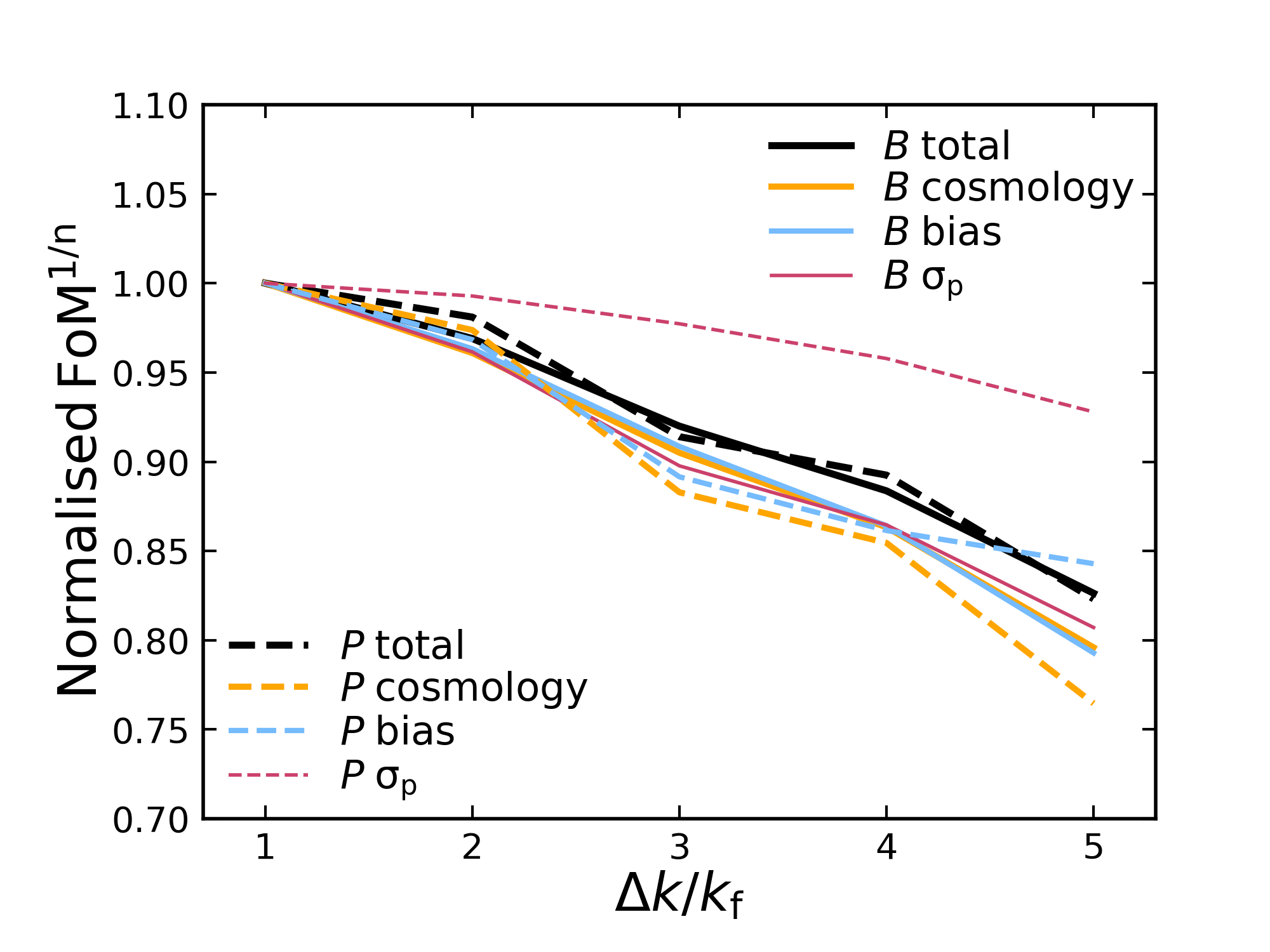}
  \caption{Typical information loss per model parameter as a function
of the bin size $\Delta k$. 
Shown is the function
FoM$^{1/n}(\Delta k)$ normalized to one at
$\Delta k=k_{\rm f}$ (the value we used in Section~\ref{results}).
Results for the power spectrum and the bispectrum measured from a \textit{Euclid}-like survey 
at $0.95<z<1.05$
are shown with dashed and solid lines, respectively. Model parameters are grouped in different sectors as indicated by the labels.
The figure refers to the
$\Lambda$CDM model.
}  
  \label{fig:fom_delta_k} 
\end{figure}

So far, we have presented results
obtained using narrow wavenumber bins with $\Delta k = k_{\rm{f}}$. This choice
is motivated by the trade-off between
minimizing information loss and taking into account the effect of the window function of the survey.
However, it is difficult to imagine
that such narrow bins will be ever used in actual observational studies. This is mainly because the large dimensionality of the data makes the estimation of covariance matrices
prohibitive, at least when it is
done using a large number of mock
galaxy catalogues.
Here, we quantify the influence of 
the bin size $\Delta k$ on
the forecast results. 
As a measure of information content,
we generalize the definition of FoM
given in equation~(\ref{eq:FoM})
and write
\be
\label{eq:genFoM}
\textrm{FoM}=\frac{1}{ \sqrt{\textrm{det}\, \textrm{Cov}(p_1,\dots,p_n)}}\;,
\ee
where $(p_1,\dots,p_n)$ denotes
the set of model parameters that belong to a given sector (e.g. 
`cosmology', `bias', etc.).
Note that
the quantity
${\mathrm{FoM}}^{1/n}$
gives an effective error estimate
for a single parameter.
In Fig.~\ref{fig:fom_delta_k},
we illustrate how the forecast constraints
from the analysis of the power spectrum and the bispectrum
degrade as the size of $\Delta k$ increases.
Shown is the ratio $\mathrm{FoM}^{1/n}(\Delta k)/ \mathrm{FoM}^{1/n}(\Delta k=k_{\rm f})$ that provides
an indication of the mean information loss per model parameter and allows
us to easily compare results obtained for different sectors.
In all cases, the deterioration of the constraints with increasing $\Delta k$ is noticeable.
For instance,
using $\Delta k=5 k_{\rm f}$ typically leads to error bars on the model parameters that are 20 per cent larger than in our reference case.
Note that the recent analysis of the bispectrum monopole from the BOSS DR12 
CMASS sample \citep{Gil_Marin_fs8} adopts $\Delta k=6 k_{\rm f}$ due to the limited number of mock catalogues available to estimate the covariance matrix.
It is only by 
compressing the data vector
with the Karhunen-Lo\`eve transform that 
\citet{Gualdi+2018_1} could employ
thinner $k$-bins down to $\Delta k=2 k_{\rm f}$.

%%%%%%%%%%%%%%%%%
%%%%%%%%%%%%%%%%%%%%%%%%%%%%%%%%%%%%%%%%%%%%%%%%%%%%%%%%%%%%%%%%%%%%%%%%%%
\subsection{Dependence on $k_{\rm{max}}$}
\begin{figure} 
	\includegraphics[width=\columnwidth]{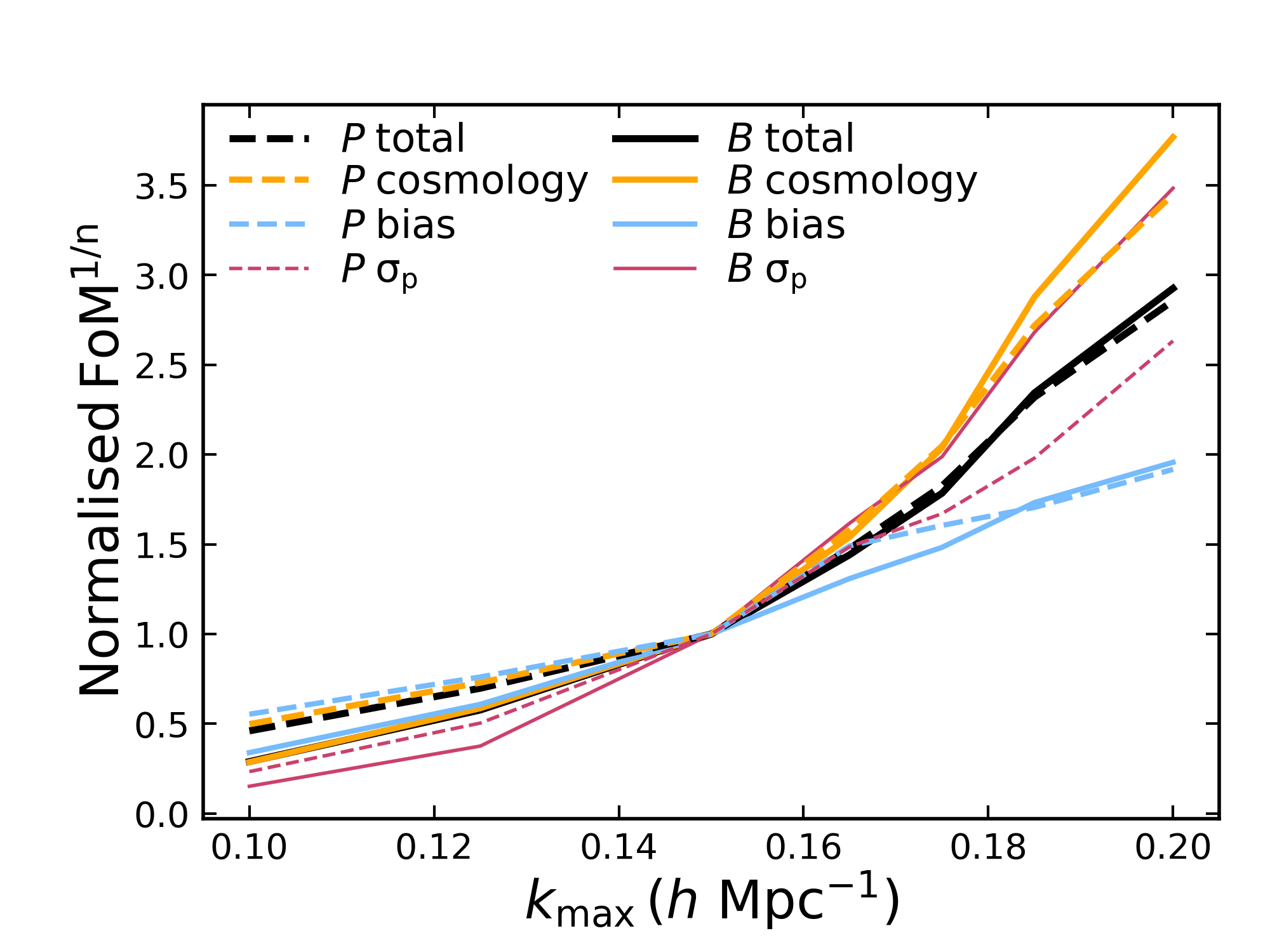}
  \caption{As in Fig.~\ref{fig:fom_delta_k} but
  as a function of $k_{\rm max}$.
  }
  \label{fig:fom_kmax} 
\end{figure}

The results presented in Section~\ref{results} have been
obtained considering all Fourier modes with $k<k_{\rm{max}}=0.15\, h\ \rm{Mpc}^{-1}$.
This choice was dictated primarily by
theoretical limitations. In fact, it is challenging to
develop models for the galaxy bispectrum in redshift space that
are sufficiently accurate on smaller scales. However, it is difficult to draw a precise line that marks where models lose their predictive power.
For this reason, here we explore how
the Fisher-matrix forecast depends
on the choice of $k_{\rm{max}}$.
An alternative approach would be
to include `theoretical errors' in the likelihood and extend the analysis to large wavenumbers
\citep{Baldauf+2016}.
Though, this would force us to always deal with impractically large covariance matrices and, also, assumptions would have to be made in order to estimate the size of the theoretical errors for the bispectrum in redshift space.
For these reasons, we prefer to use the more traditional
method of varying $k_{\rm{max}}$.
Our results are presented in Fig.~\ref{fig:fom_kmax}.
For the cosmology sector,
the quantity FoM$^{1/n}$ scales as $k_{\rm max}^\alpha$ with
$\alpha\simeq 2.7$ for the power spectrum and $\alpha \simeq 3.6$ for
the bispectrum. 
If these scaling properties can be
extrapolated beyond $0.2\,h$ Mpc$^{-1}$, our results imply that  
the bispectrum will achieve the
same constraining power as the power spectrum for $k_{\rm max}\sim 0.43\,h$ Mpc$^{-1}$.

%%%%%%%%%%%%%%%%%%%%%%%%%%%%
%%%%%%%%%%%%%%%%%%%%%%%%%%%%%%%%%%%%%%%%%%%%%%%%%%%%%%%%%%%%%%%%%%%%%%%%

\subsection{Binning of
triangle orientations}
\label{angbintest}
\begin{figure} 	\includegraphics[width=\columnwidth]{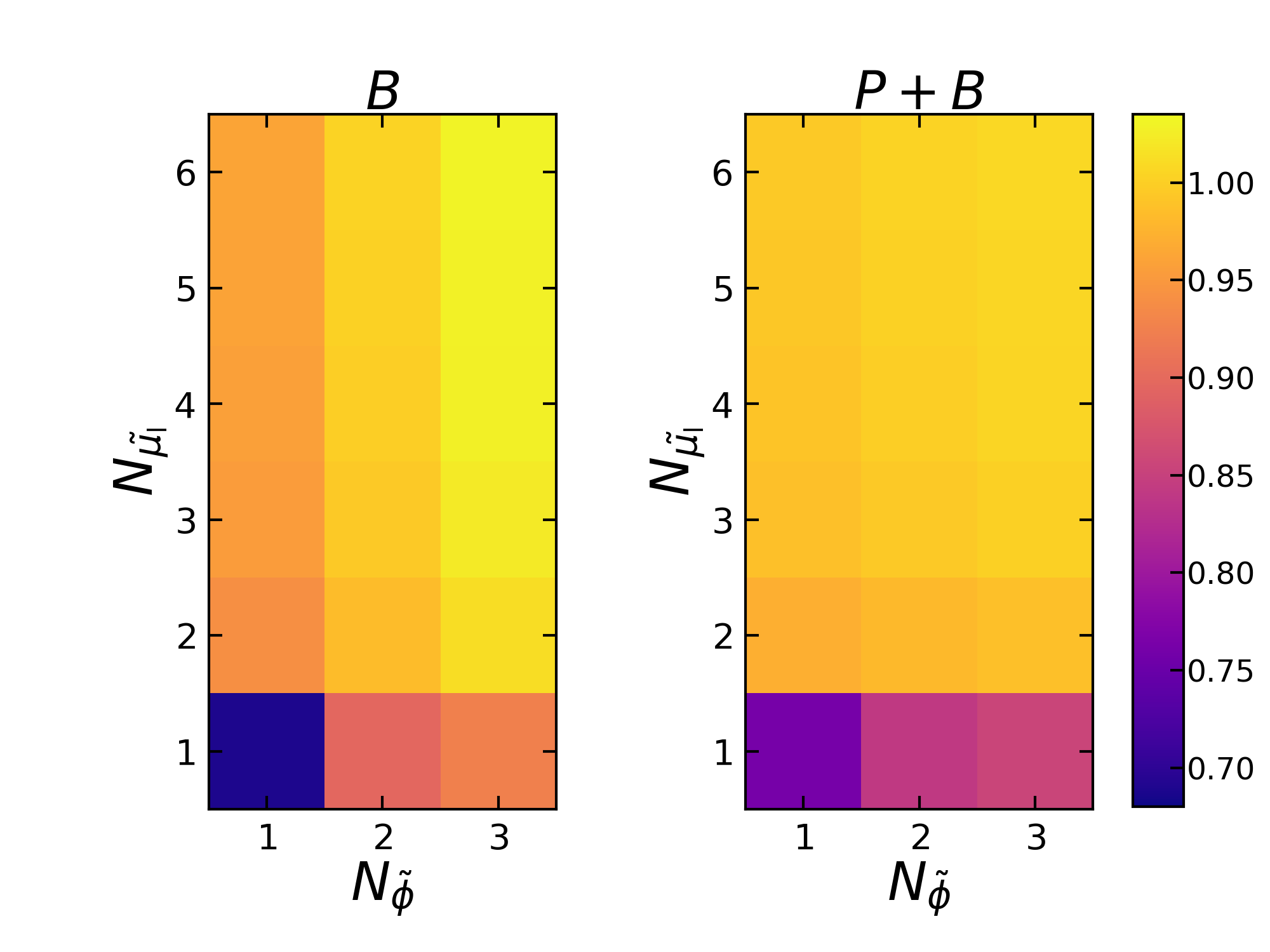}
  \caption{ As in Fig.~\ref{fig:fom_delta_k} but
  as a function of the number of bins
  used to describe the orientation
  of the triangular configuration of wavevectors with respect to the line of sight.
  Here, $N_{\tilde{\phi}}$ and
  $N_{\tilde{\mu}_{\rm l}}$
  denote the number of bins in the azimuthal angle $\tilde{\phi}$ and
  in the cosine of the polar angle $\tilde{\theta}$ (measured with respect to the longest wavevector), respectively. 
Shown is the quantity FoM$^{1/n}$ evaluated for a generic
$(N_{\tilde{\phi}},N_{\tilde{\mu}_{\rm l}})$ pair divided by the value
it assumes for our reference case
$N_{\tilde{\phi}}=2$ and
$N_{\tilde{\mu}_{\rm l}}=4$.
}
  \label{fig:NN} 
\end{figure}

In Fig.~\ref{fig:NN}, we investigate how the quantity FoM$^{1/n}$ depends on the number
of bins used to describe the
orientation of the triangular configurations for the bispectrum with respect to the line of sight.
For simplicity,
we only show results for the complete fit including all cosmological and nuisance parameters (that we
labelled `total' in Figs.~\ref{fig:fom_delta_k} and \ref{fig:fom_kmax}) as the individual plots for the different sectors all appear very similar.
The first important thing to mention is that just considering the monopole of the bispectrum in redshift space 
(i.e. $N_{\tilde{\phi}}=N_{\tilde{\mu}_{\rm l}}=1$)
leads to a non-negligible loss of information. In this case, individual parameter constraints degrade, on average, by $\sim30$ per cent with respect to our reference case ($N_{\tilde{\phi}}=2$, $N_{\tilde{\mu}_{\rm l}}=4$).
Taking into account the lowest-order non-vanishing multipoles
with $m=0$ 
(i.e. setting $N_{\tilde{\phi}}=1$ but $N_{\tilde{\mu}_{\rm l}}>1$)
is already enough to recover most
of the lost information
\citep[see also][]{Gagrani-Samushia}. However, it is necessary to also consider
the variation of the bispectrum with respect to the azimuthal angle
in order to further shrink the parameter constraints by 7 (for $B$) and 1.5 (for $P$ and $B$ combined)
per cent. Note that our reference case represents a good compromise 
between minimizing the number of bins
and keeping most of the information contained in the data.

\subsection{Shot-noise subtraction}
In line with previous theoretical work \citep[e.g.][]{SefusattiBisp,Song15,Gualdi+2018_1},
the results presented in 
Section \ref{results} 
quantify the
cosmological dependence of the actual galaxy-clustering signal
and thus assume that
we can perfectly subtract the systematic shot-noise contributions
to the power spectrum and the bispectrum. 
In a real survey, however, 
the mean galaxy density and the shot-noise corrections can only be estimated with some uncertainty
\citep[e.g. by using the selection function and the mask,][]{Feldman-Kaiser-Peacock-1994,Scoccimarro2000, Scoccimarro2}. 
Moreover, it is reasonable to expect
that shot noise is not exactly Poissonian as evidenced by the analysis of mock catalogues based
on $N$-body simulations \citep[e.g.][]{MoWhite1996, Hamaus10,Baldauf13}.
Therefore, various approaches have been taken in the literature to generalize
 equations (\ref{SNP}) and (\ref{SNB}).
For instance, in their analysis
of the BOSS survey,
\citet{S2_GilMarin} rescale the shot-noise terms $P'_{\rm shot}$ and $B_{\rm shot}$ by the same constant factor that is then fit to the data.
Similarly,
\citet{Schmittfull+2015} 
use two scale-independent factors 
to correct $P'_{\rm shot}$ and $B_{\rm shot}$ in order to fit the bispectrum
of dark-matter haloes extracted from
$N$-body simulations.
This phenomenological approach can 
be motivated by writing a more general bias expansion that
includes stochastic contributions 
\citep[e.g.][and references therein]{Dekel+Lahav-99, Matsubara99, Angulo+2015, Senatore-2015, Desj-Jeong-Schmidt2018}.
In this case, the term
$\epsilon(\bx)+\epsilon_1(\bx)\delta(\bx)$ should be added to the right-hand side of equation~(\ref{bias}).
Here, $\epsilon$ denotes the leading stochastic contribution to the bias relation while $\epsilon_1$ is the random part of the linear bias. 
By definition, both these fields have zero mean. 
Within these assumptions,
it is straightforward to show that
the power spectrum and the bispectrum
of $\epsilon$ replace
$P_{\rm shot}$ and $B_{\rm shot}$ in
equations~(\ref{SNP}) and (\ref{SNB}), respectively.
On the other hand, the cross-spectrum between $\epsilon$ and $\epsilon_1$
takes the place of $P'_{\rm shot}$ in equation~(\ref{SNB}).
A popular strategy is to assume
that, on large scales, 
these terms are approximately constant and somewhat close to the predictions of Poisson sampling.
In this Section,
we explore the consequences of
considering  
$P_{\rm shot}$, $P'_{\rm shot}$, 
and $B_{\rm shot}$ as three additional free parameters (using
the fiducial values $\ngal^{-1}, \ngal^{-1}$, and $\ngal^{-2}$, respectively).
The same approach 
has been adopted by
\citet{Karagiannis+2018} to study the constraining power of the
galaxy bispectrum on primordial non-Gaussianity.

For the power spectrum, we find that fitting the amplitude of the additional white noise term, $P_{\rm shot}$, 
worsens the constraints on all cosmological parameters by between 21 and 32 per cent (the worst case being for $n_{\rm s}$) while
basically leaves the errors
on $b_1$ and $\sigma_{\rm p}$ unchanged.

A quick look at
equation~(\ref{SNB}) shows that the situation is more complex for the bispectrum as
the shot-noise contribution also contains a scale-dependent part that is proportional to the sum of three power spectra.
For this reason, 
if we repeat the forecast presented in Section~\ref{results} by taking into account shot noise and assuming that (i) equation (\ref{SNB}) exactly applies and (ii) we perfectly know $\bar{n}$, then most of the constraints on the fit parameters improve. The largest upgrades take place for $\Omega_{\rm cdm}$ (66 per cent), $n_{\rm s}$ (47 per cent), and $A$ (44 per cent) while the smallest one
applies to $\sigma_{\rm p}$ (30 per cent).
Only the marginalized constraints in the non-linear bias parameters get slightly worse (by 7 per cent for $b_2$ and by 3.5 per cent for $b_{s^2}$).

We can now relax assumptions (i) and (ii) above by replacing $\ngal^{-1}$ and
$\ngal^{-2}$ in equation~(\ref{SNB}) with two independent free parameters,
$P'_{{\rm shot}}$ and $B_{{\rm shot}}$, that are then fit to the data including shot noise. 
After marginalizing the posteriors
over $P'_{{\rm shot}}$ and $B_{{\rm shot}}$, we find that the constraints
on $b_{\rm s^2}$ and $\sigma_{\rm p}$ worsen by nearly 50 and 30 per cent, respectively, compared with our reference case while those on the cosmological parameters
improve nearly as much as in the example discussed in the previous paragraph.

Similar outcomes are found when we combine the power spectrum and
the bispectrum: the constraint on $\Omega_{\rm cdm}$ improves by 55 per cent with respect to the corresponding reference case in Table~\ref{tab:cdm2}, those on $n_{\rm s}$ and $A$ by nearly
30 per cent, while the error on $b_2$ increases by a factor of 3. This happens because $b_2$ is degenerate with $B_{\rm shot}$.
Using the \textit{Planck} prior mitigates the differences. 
In this case, the uncertainties for all fit parameters deteriorate by less than 30--40 per cent with respect to the corresponding reference case.

The tests presented above have been performed at $z=1$ where the systematic shot-noise contribution is $\sim10$ per cent of the actual clustering signal for both $P$ and $B$. Of course, the impact of shot noise becomes more marked at higher redshifts were the number density of galaxies drops significantly. At $z\sim 2$,
for instance, shot noise is comparable with the clustering signal.

Based on these results, we conclude that the treatment of shot noise in pure clustering studies (i.e. without external priors)
has an impact on the resulting cosmological constraints and can alter them significantly. 
The tests performed here
also suggests that
our main analysis might be conservative for parameters
like $\Omega_{\rm cdm}$, $n_{\rm s}$, and $A$. 

\subsection{Treatment of galaxy bias}
In our main analysis, we have
used 3 bias parameters per redshift bin (for a total of 42) and fit them independently to the data. This is the safest approach as it does not rely
on any other assumption than the bias expansion given in equation (\ref{bias}).
However, it is reasonable to expect
that the bias parameters change smoothly with redshift. In this case,
it makes sense to approximate each of them with with a simple fitting function that captures their variation.
We consider here a quadratic function of redshift for each bias
coefficient. This reduces the number of nuisance parameters with respect to our standard treatment from 42 to 9.
Our results show that implementing this simplified procedure does not give any practical advantage as the errors on the cosmological parameters basically remain unchanged with respect to our standard treatment.

\section{Summary and conclusions}
\label{conclusions}
Galaxy clustering is a powerful
cosmological probe. 
Two-point statistics in configuration
and Fourier space are routinely used
to constrain models for our Universe. 
The question addressed in this paper
is whether the galaxy bispectrum
in redshift space 
contains additional information
about the cosmological parameters.

The literature about
the galaxy bispectrum
mostly focuses either on the real-space statistic or on its 
redshift-space monopole. 
For this reason, in Section \ref{binningstrat},
we first illustrate the phenomenology
of RSD for
the bispectrum and explore different
parameterizations for the spatial orientation
of the triangles of wavevectors
with respect to the line of sight.
We then generalize the expressions found in the literature for the
covariance matrix of bispectrum
estimates and, in particular, for their cross-covariance
with measurements of the power spectrum -- see equation (\ref{eq:CPB}).

We use the Fisher information
matrix to forecast constraints on 
a large number of cosmological and nuisance parameters from future measurements of
the galaxy bispectrum and the power spectrum in redshift space. We consider flat FLRW models dominated by dark energy and CDM with Gaussian primordial perturbations.
As an example of the forthcoming
generation of experiments, we 
adopt the specifications of
a \textit{Euclid}-like galaxy redshift survey
(Table~\ref{tab:Euclid}).
In our principal analysis, we only
consider wavenumbers with $k\leq k_{\rm max}=0.15\,h$ Mpc$^{-1}$ that define mildly non-linear scales on which fluctuations in the galaxy density can be treated perturbatively.
Within this range,
it should thus be possible to build robust models for the galaxy power spectrum and bispectrum.
The main conclusions of our work are
as follows:

(i) The galaxy bispectrum and the power spectrum in redshift space set constraints of similar strength on the cosmological parameters (Table~\ref{tab:cdm2}). Therefore the bispectrum can be used as a consistency check for power-spectrum studies.

(ii) Posterior correlations between the model parameters derived from the bispectrum and the power spectrum are, in most cases, very similar (Figs.~\ref{fig:el_cdm}--\ref{fig:el_w0wacdm}). For this reason, combining the two probes only moderately improves
the cosmological constraints with respect to
considering them individually.

(iii) For instance, considering both statistics together partially breaks the degeneracies between the linear bias coefficients and the galaxy-clustering amplitudes in all redshift bins. In consequence,
the FoM for the dark-energy
parameters $w_0$ and $w_a$ 
improves by a factor of 2.6 with respect to only using the power spectrum. 

(iv) This advantage, however, vanishes
once priors based on the results of the \textit{Planck} mission and of current clustering studies are included in the analysis. In this case, combining the power spectrum with the bispectrum does not give any appreciable benefit other than precisely determining the parameters that describe galaxy bias.

(v) For wavenumbers $k<0.15 \,h$ Mpc$^{-1}$, the cross-covariance between the power spectrum and the bispectrum has a small influence on parameter estimation (Table~ \ref{tab:cdm2})
and may be safely neglected to first approximation.

(vi) Taking broad bins for the legs of the triangles of wavevectors 
leads to some information loss for the bispectrum (Fig.~\ref{fig:fom_delta_k}). For instance, using $\Delta k=5 k_{\rm f}$ gives cosmological constraints which are suboptimal by 20 per cent.

(vii) Since the number of bins in the triangular configurations for the bispectrum grows more rapidly with the maximum 
wavenumber than
the number of bands in the power spectrum,
the relative importance of the two probes
strongly depends on the value of $k_{\rm max}$ that is considered (Fig.~\ref{fig:fom_kmax}).
We find that, for $k_{\rm max}=0.15 \,h$ Mpc$^{-1}$, the
power spectrum provides slightly tighter constraints than the bispectrum on most parameters.
However, our results also suggest that the bispectrum becomes the leading probe
if the analysis is extended beyond $k_{\rm max}\simeq 0.43 \,h$ Mpc$^{-1}$ (assuming that an accurate theoretical model is available at such wavenumbers).

(viii) Redshift-space distortions contain precious information about the cosmological parameters. 
Just considering the monopole moment of the bispectrum 
leads to a non-negligible loss of information. Individual error bars for the fit parameters typically grow by 50 per cent (Fig.~\ref{fig:NN}).
Taking into account the lowest-order non-vanishing multipoles with $m=0$ recovers most of the lost information. Considering also variations of the bispectrum with the azimuthal angle further reduces the error bars by a few up to 10 per cent.

(ix) The way shot noise is handled
in the clustering analysis influences the cosmological results
(especially for $\Omega_{\rm cdm}, n_{\rm s}$ and
$A$) as well as the non-linear bias parameter $b_2$.
However, this dependence is significantly reduced by also considering CMB-based priors.

(x) Using a smooth function of redshift to describe the evolution of the bias coefficients does not lead to any practical advantage with respect to fitting individual parameters for every redshift bin.

\section*{Acknowledgements}
We thank Emiliano Sefusatti for useful discussions and the anonymous reviewer for suggestions that helped improving the presentation of our results.
We acknowledge financial support by the Deutsche Forschungsgemeinschaft through the Transregio 33 `The Dark Universe'. VY was also partly supported through a research contract from the International Max Planck Research School (IMPRS) for Astronomy and Astrophysics at the Universities of Bonn and Cologne and partially supported by the Bonn-Cologne Graduate School for Physics and Astronomy.
%%%%%%%%%%%%%%%%%%%%%%%%%%%%%%%%%%%%%%%%%%%%%%%%%%



\bibliographystyle{mnras}
\bibliography{yankelevich} % if your bibtex file is called example.bib

%%%%%%%%%%%%%%%%%%%%%%%%%%%%%%%%%%%%%%%%%%%%%%%%%%
\appendix 
\section{Coordinate systems}
\label{App:RSDs}
\begin{figure}
\includegraphics[width=\columnwidth]{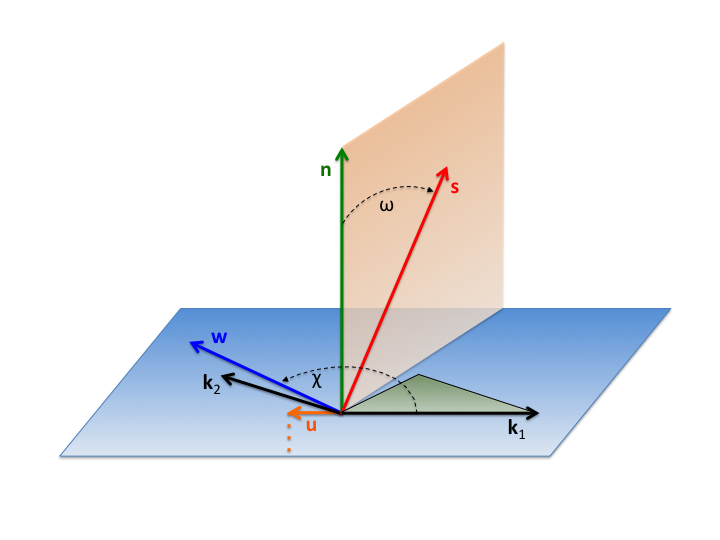}
  \caption{Schematic showing the definition of the angles $(\omega,\chi)$.} 
  \label{fig:coords1} 
\end{figure}
We introduce here two different
coordinate systems 
in order to parameterize the relative orientation between a triangle of wavevectors and the line of sight.

\subsection{Using the triangle's normal as the polar axis}
Let us consider a triangle of sides $\bk_1$, $\bk_2$ and $\bk_3$ such that $\bk_1+\bk_2+\bk_3=0$.
The triangle lies on a plane whose normal vector is parallel to $\bn=\bk_1\times \bk_2$.
The orientation of the unit vector $\hat{\bn}=\bn/||\bn ||$ with respect to the line-of-sight direction $\hat{\bs}$ can be described in terms of a single rotation around the axis $\bw=\hat{\bn}\times \hat{\bs}$ (see Fig.~\ref{fig:coords1}).
We want to build a right-handed orthonormal basis starting from $\hat{\bs}$ and $\hat{\bw}$. For the third element of the basis we pick a unit vector $\hat{\bu}$ parallel to $\hat{\bs}\times\hat{\bw}=\hat{\bn}-(\hat{\bn}\cdot \hat{\bs})\,\hat{\bs}$, i.e. $\hat{\bu}=(\hat{\bn}-(\hat{\bn}\cdot \hat{\bs})\,\hat{\bs})/\sqrt{1-(\hat{\bn} \cdot \hat{\bs})^2}$.
In the base $\hat{\bw}$, $\hat{\bu}$, $\hat{\bs}$, the rotation from $\hat{\bn}$ to $\hat{\bs}$ is described by the matrix,
\begin{equation} 
\mathbfss{R}=
\begin{pmatrix}
1 & 0 & 0 \\
0 & \hat{\bn}\cdot\hat{\bs} & -||\hat{\bn}\times \hat{\bs} || \\
0 & ||\hat{\bn}\times \hat{\bs}|| & \hat{\bn}\cdot\hat{\bs}
\end{pmatrix}\;.
\end{equation}
In fact, $\hat{\bn}$ is a column vector with coordinates 
\begin{equation}
(\hat{\bn}\cdot \hat{\bw}, \hat{\bn}\cdot \hat{\bu}, \hat{\bn}\cdot \hat{\bs})=(0,\sqrt{1-(\hat{\bn}\cdot\hat{\bs})^2},\hat{\bn}\cdot\hat{\bs})
\end{equation}
 and applying the rotation to it one gets $(0,0,1)$.
This is a rotation by an angle $0\leq\omega<\pi$ such that $\cos\omega= \hat{\bn}\cdot\hat{\bs}$ and $\sin\omega=||\hat{\bn}\times \hat{\bs} ||=||\bw||$
(note than $\sin \omega\geq0$).
This completely describes the relative orientation of the plane of the triangle with respect to the line of sight.

Now, we only need to describe the orientation of the triangle on its plane.
Note that, being perpendicular to $\hat{\bn}$,  the basis element $\hat{\bw}$ lies on the plane of the triangle.
It is thus convenient to measure the orientation of the triangle in its plane by looking at the orientation of, say, $\bk_1$ with respect to $\hat{\bw}$. In order to quantify this, we  introduce the angle $\chi$ ($0\leq \chi<2\pi$)
such that $\hat{\bk}_1\cdot \hat{\bw}=\cos \chi$ and $(\hat{\bk}_1\times \hat{\bw})\cdot\hat{\bn}=\sin \chi$.
It is worth stressing that $\hat{\bk}_1\times \hat{\bw}= \hat{\bk}_1\times (\hat{\bn}\times \hat{\bs})/||\bw||=
(\hat{\bk}_1\cdot \hat{\bs})\,\hat{\bn}/||\bw||$ and $\sin \chi=(\hat{\bk}_1\cdot \hat{\bs})/||\bw||=\mu_1/\sin\omega$.
The angle $\chi$ denotes the %(shortest) 
rotation angle around $\hat{\bn}$ from $\hat{\bk}_1$ to $\hat{\bw}$.

\begin{figure}
\includegraphics[width=\columnwidth]{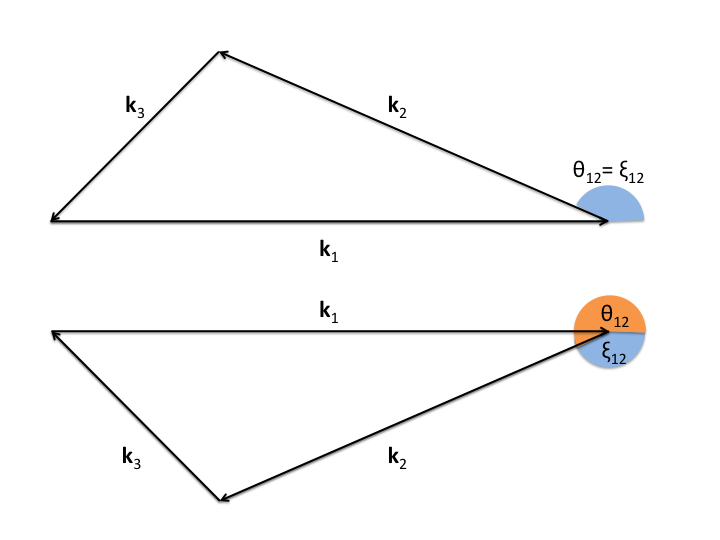}
  \caption{Definition of the angles $\theta_{12}$ and $\xi_{12}$ for two triangles with the same shape but opposite handedness.} 
  \label{fig:angle12} 
\end{figure}

Let us now reverse the problem and determine the line-of-sight components of $\bk_1, \bk_2, \bk_3$ for given $\omega$ and $\chi$. 
The shape and the handedness of the triangle matter.
A common choice is to parameterize the relative orientation of $\bk_1$ and $\bk_2$ in terms of the angle $\theta_{12}$ such that 
$\hat{\bk}_2\cdot \hat{\bk}_1=\cos \theta_{12}$ and $||\bn||=||\bk_1\times \bk_2||=|\sin \theta_{12}|$.
In principle, $0\leq\theta_{12}<2\pi$ and, for a fixed shape, triangles with $\theta_{12}$ and $2\pi-\theta_{12}$ have opposite handedness (see Fig.~\ref{fig:angle12}).  However, $\hat{\bn}$, $\hat{\bw}$ and $\hat{\bu}$ flip sign when the handedness is switched.
It is thus much more convenient 
to express the shape of the triangle in terms of a rotation angle around $\hat{\bn}$ and always use an angle $\xi_{12}$ such that $0\leq \xi_{12}<\pi$ and $\sin \xi_{12}\geq 0$. 
In words, $\xi_{12}=\arccos (\hat{\bk}_1\cdot\hat{\bk}_2)$ is the (shortest) rotation angle around $\hat{\bn}$ from $\hat{\bk}_1$ to $\hat{\bk}_2$.
Triangles with the same shape but opposite handedness have identical $\xi_{12}$.

%%%%%
%
We recall that,  using the vector basis we have introduced above, $\hat{\bn}=(0,\sin \omega, \cos\omega)$ and $\hat{\bs}=(0,0,1)$, so that
$\bw=(\sin \omega,0,0)$ and $\bu=(0,\sin\omega,0)$.
From the definitions $\hat{\bk}_1\cdot \hat{\bw}=\cos \chi$ and $(\hat{\bk}_1\times \hat{\bw})\cdot\hat{\bn}=\sin \chi$, it follows that
\begin{equation}
\bk_1=k_1\,(\cos \chi, -\cos\omega \sin \chi, \sin \omega \sin \chi)\;.
\end{equation}
Since the vector $\hat{\bk}_2$ corresponds to a rotation of $\hat{\bk}_1$ by an angle $\xi_{12}$ around $\hat{\bn}$ while
$\hat{\bw}$ is rotated from $\hat{\bk}_1$ by an angle $\chi$ around $\hat{\bn}$, it follows that 
\begin{equation}
\bk_2=k_2\,(\cos (\chi-\xi_{12}), -\cos\omega \sin (\chi-\xi_{12}), \sin \omega \sin (\chi-\xi_{12}))\;.
\end{equation}
This univocally fixes the RSD:
\begin{eqnarray}
\mu_1&=& \sin \omega \sin\chi\;, \label{mymu1}\\ 
\mu_2&=& \sin\omega \sin (\chi-\xi_{12})\;. \label{mymu2} 
\end{eqnarray}
\begin{figure}
\includegraphics[width=\columnwidth]{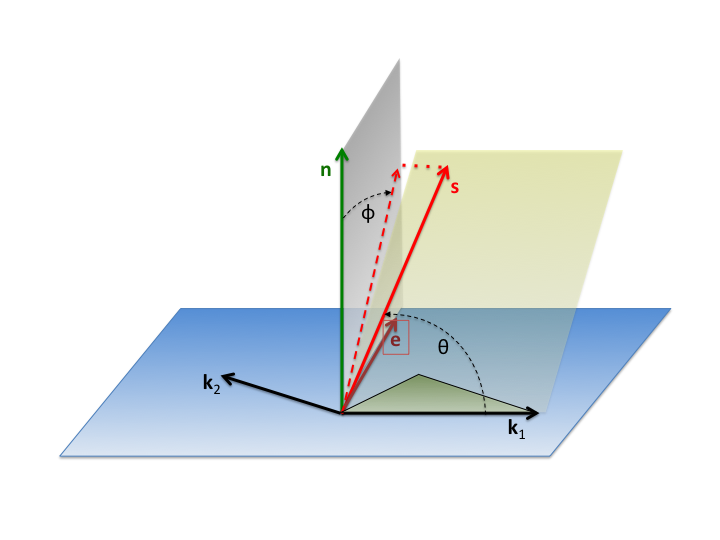}
  \caption{Schematic showing the definition of the angles $(\theta,\phi)$.} 
  \label{fig:coords2} 
\end{figure}
\subsection{Using $\bk_1$ as the polar axis}
Scoccimarro et al. (1999) use a different parameterization in terms of the polar angle $0\leq\theta<\pi$ and the azimuthal angle $0\leq \phi<2\pi$ that define the orientation of $\hat{\bs}$ with respect to $\hat{\bk}_1$ (see Fig.~\ref{fig:coords2}). 
In order to link this approach to our previous discussion, let us build
a right-handed orthonormal basis by complementing $\hat{\bn}$ and $\hat{\bk}_1$ with another unit vector $\hat{\bee}$ lying
in the plane of the triangle -- i.e, $\hat{\bee}$ is the unit vector of $\bee=\bn\times \bk_1=k_1^2\,\bk_2-(\bk_1\cdot\bk_2)\,\bk_1$ or
$\hat{\bee}=[\hat{\bk}_2-(\hat{\bk}_1\cdot\hat{\bk}_2)\,\hat{\bk}_1]/\sqrt{1-(\hat{\bk}_1\cdot \hat{\bk}_2)^2}$.
In the basis $(\hat{\bk}_1,\hat{\bee},\hat{\bn})$, $\hat{\bk}_2$ is a column vector of coordinates
$(\cos \xi_{12},\sin \xi_{12},0)$ - note once again that both $\hat{\bn}$  and $\hat\bee$ flip sign if the handedness of the triangle is changed and this is why we can use $\xi_{12}$ instead of $\theta_{12}$.
For the azimuth $\phi$, we use the angle between $\hat{\bn}$ and the projection of $\hat{\bs}$ on the plane defined by $\hat{\bn}$ and $\hat\bee$. This means that $\cos\phi=0$ (i.e. $\phi=\pi/2$ or $3\pi/2$) whenever $\hat{\bs}$ lies on the plane of triangle. Given all this, in the basis $(\hat{\bk}_1,\hat{\bee},\hat{\bn})$,
$\hat{\bs}$ is the column vector of coordinates
$(\cos \theta,\sin\theta\,\sin\phi,\sin\theta\,\cos\phi)$ so that
\begin{eqnarray}
\mu_1&=&\cos \theta\;, \label{scomu1}\\
\mu_2&=&\cos \theta \cos\xi_{12}+ \sin \theta \sin\phi \sin \xi_{12}\;. \label{scomu2}
\end{eqnarray}

For generic vectors $\bk_1, \bk_2$ and $\hat{\bs}$ defined in an arbitrary basis (e.g. a Fourier grid used to measure the bispectrum in
a numerical simulation or for a galaxy catalogue), the angles 
$\theta$ and $\phi$ can be determined as follows. The polar angle is simply given by $\theta=\arccos(\hat{\bk}_1\cdot \hat{\bs})$.
For the azimuth, instead, it is convenient to introduce the vector $\bs_\perp=\hat{\bs}-(\hat{\bs}\cdot \hat{\bk}_1)\, \hat{\bk}_1$ (which gives the
component of $\hat{\bs}$ perpendicular to $\bk_1$) and calculate the real numbers $\cos\phi=\hat{\bs}_\perp\cdot\hat{\bn}=
(\hat{\bs}\cdot \hat{\bn})/||\bs_\perp|| = \sigma_n$ and
$\sin\phi=\hat{\bs}_\perp\cdot\hat{\bee}=(\hat{\bs}\cdot\hat{\bee})/||\bs_\perp||=\sigma_e$.
If $\sin\phi>0$, then $\phi=\arccos(\sigma_n)$ while, if $\sin\phi<0$, $\phi=2\pi-\arccos(\sigma_n)$.

\subsection{Matching the different coordinate systems}

Starting from the expressions for $\bk_1$, $\bk_2$ and $\bs$ in the $(\theta,\phi)$ coordinates and applying the definitions of the
angles $\omega$ and $\chi$, one obtains:
\begin{eqnarray}
\cos\omega&=&\sin\theta \cos\phi\;,\\
\sin\omega&=&\sqrt{1-\sin^2\theta \cos^2\phi}\;,\\
\cos\chi&=&\frac{-\sin \theta \sin\phi}{\sqrt{1-\sin^2\theta \cos^2\phi}}\;,\\
\sin\chi&=&\frac{\cos \theta}{\sqrt{1-\sin^2\theta \cos^2\phi}}\;.
\end{eqnarray}

Vice versa, starting from the expressions in terms of $(\omega,\chi)$, one derives:
\begin{eqnarray}
\cos\theta&=&\sin\omega \sin\chi\;,\\
\sin\theta&=&\sqrt{1-\sin^2\omega \sin^2\chi}\;,\\
\cos\phi&=& \frac{\cos\omega}{\sqrt{1-\sin^2\omega \sin^2\chi}}\;,\\
\sin\phi&=& \frac{-\sin \omega \cos\chi}{\sqrt{1-\sin^2\omega \sin^2\chi}}\;.
\end{eqnarray}

\subsection{Symmetries}
\label{rsd:symmetries}
RSD are quadratic in the $\mu_i$ and do not change if $\mu_1, \mu_2$ and $\mu_3$ change sign
simultaneously. 
In terms of the $(\theta, \phi)$ variables, this means that the galaxy bispectrum in redshift space is symmetric with respect to the transformation
\begin{equation}
\begin{cases}
\theta \to \pi-\theta\;, \\
\phi \to 2\pi-\phi\;. 
\end{cases}
\end{equation}
In fact, by considering equations (\ref{scomu1}) and (\ref{scomu2}), 
one can easily prove that this transformation changes sign to $\mu_1, \mu_2$ and $\mu_3$.
This means that the $(\theta, \phi)$ variables are somewhat redundant and
not all the
parameter space they cover is necessary to describe
the RSD.
It is thus appealing to seek for 
new angular coordinates that make
the necessary region more compact and do not present duplications.
In fact, this helps reduce the number of bins needed to represent all possible configurations.
For instance, we can halve the
size of parameter space by
introducing a new set of coordinates $(\tilde{\theta},\phi')$ such
that 
$\tilde{\theta}=\min(\theta,\pi-\theta)$ and
\begin{equation}
\phi'=\begin{cases}
\phi\;, & \text{if } \theta<\pi/2\;, \\
2\pi-\phi\;, & \text{otherwise.}
\end{cases}
\end{equation}
In this case,  
$0\leq\tilde{\theta}<\pi/2$
(or $0<\tilde{\mu}=\cos\tilde{\theta}\leq1$)
and $0\leq\phi'<2\pi$.
However, RSD
possess still another symmetry
deriving from the fact that they
only depend on $\sin \phi$
(and, equivalently, on $\sin\phi'$).
Since, $\sin(\pi-x)=\sin x$,
we can further halve the area 
of parameter space by introducing
the variable $\pi/2\leq\tilde{\phi}<3\pi/2$ defined as follows
\begin{equation}
\tilde{\phi}=\begin{cases}
\pi-\phi'\;, &\text{if }
0\leq\phi'<\pi/2\;, \\
\phi'\;, &\text{if }
\pi/2\leq\phi'<3\pi/2\;,\\
3\pi-\phi'\;, &\text{if } 
3\pi/2\leq\phi'<2\pi\;.
\end{cases}
\end{equation}
The angular variables $\tilde{\theta}$ and $\tilde{\phi}$
are optimal in the sense that they
suffice to describe all possible configurations while minimizing the size of parameter space.

Similarly, we can derive optimal
variables also starting from the coordinates $(\omega,\chi)$.
Equations (\ref{mymu1}) and (\ref{mymu2}) show that $\mu_1$ and $\mu_2$ change sign if $\chi \to \pi+\chi$ while $\omega$ is left
unchanged. 
It follows that considering the variable $0\leq\tilde{\chi}<\pi$ 
defined as
\begin{equation}
\tilde{\chi}=\begin{cases}
\chi\;, &\text{if } \chi<\pi\;, \\
\chi-\pi\;, &\text{otherwise}\;.
\end{cases}
\end{equation}
is sufficient to identify the configurations with opposite signs
of $\mu_1$ and $\mu_2$.
In fact,
under the transformation $\theta \to \pi-\theta$ and  $\phi \to 2\pi-\phi$,
$\omega$ is unchanged while both $\cos \chi$ and $\sin \chi$ change sign that corresponds to the transformation $\chi \to \pi+\chi$. 
The second symmetry, in this case,
derives from the fact that the RSD only depend on $\sin \omega$. Therefore, we can further reduce the extension of parameter space by introducing the
variable 
$\tilde{\omega}=\min(\omega,\pi-\omega)$
so that $\cos \tilde\omega=|\cos \omega|$.
The set $(\tilde{\omega},\tilde{\chi})$ is optimal.

\section{Bias parameters}
\label{App:table}
In Table~\ref{tab:bias}, 
we report the forecast errors for the galaxy-bias parameters corresponding to our main analysis presented in Section~\ref{results}.
\begin{table*}
	\renewcommand{\thetable}{B\arabic{table}}
	\caption{Expected marginalized $1\sigma$ errors 
		for the galaxy bias parameters $b_1$, $b_2$ and $b_{s^2}$ 
		in the $\Lambda$CDM, $w$CDM and $w_0w_a$CDM models obtained considering a \textit{Euclid}-like  survey.
		The different rows display results obtained from the galaxy power spectrum ($P$, only for $b_1$),
		the bispectrum ($B$), and their combination ($P+B$) for 14 redshift bins centred at redshift $z$ (different columns). }
	\label{tab:bias}
	\begin{tabular}{cccccccccccccccc}
		\hline
		Probe & Param.  & 0.7 &  0.8 &  0.9 &  1.0 &  1.1 &  1.2 &  1.3 &  1.4 &  1.5 &  1.6 &  1.7 &  1.8 &  1.9 &  2.0 \\ 
		\hline
		\multicolumn{16}{|c|}{$\Lambda$CDM model} \\
		\hline
		$P$ & $b_1$ &   0.014 &   0.014 &   0.015 &   0.016 &   0.017 &   0.017 &   0.018 &   0.019 &   0.020 &   0.021 &   0.022 &   0.023 &   0.024 &   0.026 \\ 
		$B$ &  $b_1$ &  0.046 &   0.048 &   0.051 &   0.053 &   0.056 &   0.060 &   0.066 &   0.072 &   0.085 &   0.097 &   0.120 &   0.157 &   0.189 &   0.253 \\ 
		$P+B$ & $b_1$ &   0.012 &   0.012 &   0.013 &   0.013&   0.014 &   0.014 &   0.015 &   0.015 &   0.016 &   0.017 &   0.018 &   0.019 &   0.020 &   0.022 \\ 
		\hline
		$B$ & $b_2$ &  0.062 &   0.064 &   0.070 &   0.073 &   0.080 &   0.089 &   0.105 &   0.125 &   0.164 &   0.203 &   0.272 &   0.384 &   0.488 &   0.684 \\  
		$P+B$ & $b_2$ &  0.014 &   0.015 &   0.016 &   0.016&   0.017 &   0.019 &   0.023 &   0.027 &   0.035 &   0.043 &   0.057 &   0.079 &   0.099 &   0.137 \\ 
		\hline
		$B$ &  $b_{s^2}$ &  0.143 &   0.144 &   0.156 &   0.159 &   0.175 &   0.194 &   0.230 &   0.270 &   0.355 &   0.432 &   0.572 &   0.795 &   0.992 &   1.365 \\  
		$P+B$ &  $b_{s^2}$ &   0.070 &   0.071 &   0.077 &   0.079&   0.086 &   0.096 &   0.113 &   0.132 &   0.173 &   0.210 &   0.277 &   0.384 &   0.479 &   0.658 \\ 
		\hline
		\multicolumn{16}{|c|}{$w$CDM model}\\
		\hline
		$P$ & $b_1$ & 0.024 &   0.027 &   0.029 &   0.032 &   0.034 &   0.037 &   0.039 &   0.042 &   0.044 &   0.047 &   0.050 &   0.052 &   0.055 &   0.058 \\ 
		$B$ & $b_1$ &   0.071 &   0.077 &   0.084 &   0.091 &   0.099 &   0.108 &   0.117 &   0.127 &   0.142 &   0.155 &   0.176 &   0.209 &   0.239 &   0.296 \\ 
		$P+B$ & $b_1 $&  0.016 &   0.017 &   0.019 &   0.020&   0.022 &   0.023 &   0.025 &   0.027 &   0.028 &   0.030 &   0.032 &   0.034 &   0.036 &   0.038 \\ 
		\hline
		$B$ & $b_2$&   0.074 &   0.077 &   0.084 &   0.088 &   0.097 &   0.107 &   0.123 &   0.142 &   0.180 &   0.218 &   0.285 &   0.394 &   0.497 &   0.691 \\  
		$P+B$ & $b_2$ &   0.015 &   0.015 &   0.016 &   0.016&   0.018 &   0.019 &   0.023 &   0.027 &   0.035 &   0.043 &   0.057 &   0.080 &   0.100 &   0.138 \\ 
		\hline
		$B$ & $b_{s^2}$ &   0.146 &   0.147 &   0.158 &   0.160 &   0.175 &   0.194 &   0.230 &   0.270 &   0.355 &   0.433 &   0.573 &   0.795 &   0.993 &   1.366 \\ 
		$P+B$ & $b_{s^2}$ &  0.073 &   0.075 &   0.081 &   0.084&   0.092 &   0.102 &   0.119 &   0.138 &   0.178 &   0.215 &   0.282 &   0.388 &   0.483 &   0.661 \\ 
		\hline
		\multicolumn{16}{|c|}{$w_0w_a$CDM model}\\
		\hline
		$P$ & $b_1$ &  0.045 &   0.048 &   0.052 &   0.056 &   0.060 &   0.065 &   0.069 &   0.074 &   0.079 &   0.085 &   0.090 &   0.095 &   0.101 &   0.106 \\ 
		$B$ & $b_1$ &  0.090 &   0.095 &   0.101 &   0.107 &   0.115 &   0.124 &   0.134 &   0.145 &   0.160 &   0.175 &   0.196 &   0.228 &   0.259 &   0.315 \\ 
		$P+B$ & $b_1$ &  0.019 &   0.020 &   0.022 &   0.023&   0.025 &   0.027 &   0.029 &   0.031 &   0.033 &   0.035 &   0.038 &   0.040 &   0.043 &   0.045 \\ 
		\hline
		$B$ & $b_2$ &    0.087 &   0.089 &   0.093 &   0.096 &   0.103 &   0.112 &   0.127 &   0.146 &   0.183 &   0.220 &   0.287 &   0.395 &   0.498 &   0.692 \\ 
		$P+B$ & $b_2$ &  0.015 &   0.015 &   0.016 &   0.016&   0.018 &   0.019 &   0.023 &   0.027 &   0.035 &   0.044 &   0.058 &   0.080 &   0.101 &   0.138 \\   
		\hline
		$B$ & $b_{s^2}$ &  0.147 &   0.147 &   0.158 &   0.161 &   0.176 &   0.195 &   0.231 &   0.271 &   0.356 &   0.434 &   0.574 &   0.796 &   0.993 &   1.366 \\ 
		$P+B$ & $b_{s^2}$ &  0.077 &   0.078 &   0.084 &   0.087&   0.095 &   0.105 &   0.122 &   0.141 &   0.181 &   0.218 &   0.284 &   0.389 &   0.484 &   0.663 \\ 
		\hline
	\end{tabular}
\end{table*}      
%%%%%%%%%%%%%%%%%%%%%%%%%%%%%%%%%%%%%%%%%%%%%%%%%%%%%%%%%%%%%%%%%%%%%%%%%%

\end{document}